\documentclass{tlp}

\usepackage[T1]{fontenc}
\usepackage[utf8]{inputenc}

\usepackage{xcolor}
\usepackage[pdftex]{graphicx}

\usepackage{textgreek}

\usepackage{listings}

\lstnewenvironment{coqlisting}{%
  \lstset{language=Coq,fontadjust=true,columns=fullflexible,showspaces=false}
  \lstset{basicstyle=\sf\fontsize{7}{8.69}\selectfont,
    commentstyle=\it\sffamily\fontsize{7}{8}
    \color{mydarkred},%
    keywordstyle=\sffamily\color{blue}}
  \lstset{escapeinside={(*@}{@*)}}
}{%
  \vspace*{-1mm plus 1mm minus 1mm}
  \lstset{language=Coq,fontadjust=true,flexiblecolumns=true,showspaces=false}
  \lstset{basicstyle=\sf\small,commentstyle=\it\sffamily\small\color{mydarkred},keywordstyle=\bf\sffamily\color{blue}}
  \lstset{escapeinside={(*@}{@*)}}
  \lstset{commentstyle=\color{mydarkred}\it\small}
}

\newif\ifelec

\electrue
\definecolor{burntorange}{rgb}{0.8, 0.33, 0.0}

\lstdefinelanguage{SSR}{
mathescape=true,
texcl=false,
morekeywords=[1]{
Section, Module, End, Require, Import, Export, Defensive, Function, Axioms,
Variable, Variables, Parameter, Parameters, Axiom, Hypothesis, Hypotheses,
Notation, Local, Tactic, Reserved, Scope, Open, Close, Bind, Delimit,
Definition, Let, Ltac, Fixpoint, CoFixpoint, Add, Morphism, Relation,
Implicit, Identity, Types, Arguments, Unset, Contextual, Strict, Prenex, Implicits,
Inductive, CoInductive, Record, Structure, Canonical, Coercion,
Theorem, Lemma, Corollary, Proposition, Fact, Remark, Example,
Proof, Goal, Save, Qed, Defined, Hint, Resolve, Rewrite, View,
Search, Show, Print, Printing, All, Graph, Projections, inside,
outside, Locate, Maximal, Eval, Compute, Time, Check, Print, About,
Inline, Class, Instance, Context, Include, Declare},
morekeywords=[2]{forall, fun, fix, cofix, struct,
      match, with, end, as, in, return, let, if, is, then, else,
      for, of, nosimpl, where, True, False, beta, delta, zeta, iota},
morekeywords=[3]{Set, Type, Prop},
morekeywords=[4]{
         exists, exists2,
         pose, set, move, case, elim, apply, clear,
            hnf, intro, intros, generalize, rename, pattern, after,
	    destruct, induction, using, refine, inversion, injection,
            constructor,
         rewrite, congr, unlock, compute, vm_compute, native_compute,
            replace, fold, unfold, change, cutrewrite, simpl,
            cbv, lazy,
         have, suff, wlog, suffices, without, loss, nat_norm,
            assert, cut, trivial, revert, bool_congr, nat_congr,
	 symmetry, transitivity, auto, split, left, right,
         autorewrite,
       interval_intro},
morekeywords=[5]{
         by, done, exact, reflexivity, tauto, romega, omega,
         assumption, solve, contradiction, discriminate,
         ring, field, interval},
morekeywords=[6]{do, last, first, try, idtac, repeat, progress},
literate=
        {:=}{{$\mathrel{\mathop:\mathopen=}$}}2
        {=}{{$=$}}1
        {==}{{$\equiv$}}1
        {!=}{{$\not\equiv$}}1
        {<=}{{$\leq$}}1
        {>=}{{$\geq$}}1
        {<>}{{$\neq$}}1
        {->}{{$\rightarrow$}}2
        {<-}{{$\leftarrow$}}2
        {=>}{{$\Rightarrow$}}1
        {/\\}{{$\wedge$}}2
        {\\/}{{$\vee$}}2
        {<->}{{$\leftrightarrow$}}2
        {<=>}{{$\Leftrightarrow$}}2
        {forall\ }{{$\forall\,$}}1
        {exists\ }{{$\exists\,$}}1
        {negb}{{$\neg$}}1
        {~}{{$\neg$}}1
        {\\in}{{$\in$}}1
        {^-1}{{$^{-1}$}}1,
comment=[s]{(*}{*)},
showstringspaces=false,
morestring=[b]",
morestring=[d]´,
extendedchars=true,
sensitive=true,
breaklines=true,
basicstyle=\small\ttfamily,
captionpos=b,
columns=[l]fullflexible,
keepspaces=true,
identifierstyle={\ttfamily\ifelec\color{black}\fi},
keywordstyle=[1]{\ttfamily\ifelec\color{dkviolet}\fi},
keywordstyle=[2]{\ttfamily\ifelec\color{dkgreen}\fi},
keywordstyle=[3]{\ttfamily\ifelec\color{dkgreen}\fi},
keywordstyle=[4]{\ttfamily\ifelec\color{dkblue}\fi},
keywordstyle=[5]{\ttfamily\ifelec\color{red}\fi},
keywordstyle=[6]{\ttfamily\ifelec\color{dkpink}\fi},
stringstyle=\ttfamily,
commentstyle={\ttfamily\ifelec\color{firebrick}\fi},
moredelim=[is][\color{burntorange}\bfseries\ttfamily]{|*}{*|},
moredelim=*[is][\itshape\rmfamily]{/*}{*/},
moredelim=[is][\ttfamily\color{dkviolet}]{\{-}{-\}}
}

\definecolor{dkblue}{rgb}{0,0.1,0.5}
\definecolor{lightblue}{rgb}{0,0.5,0.5}
\definecolor{dkgreen}{rgb}{0,0.4,0}
\definecolor{dk2green}{rgb}{0.4,0,0}
\definecolor{dkviolet}{rgb}{0.6,0,0.8}
\definecolor{dkpink}{rgb}{0.75,0,1}
\definecolor{firebrick}{rgb}{0.69,0.13,0.13}

  {\begin{Sbox}\begin{minipage}{0.97\textwidth}%
    \begin{flushright}\textcolor{red}{\fbox{Version 1.3}}%
      \end{flushright}\noindent}%
  {\end{minipage}\end{Sbox}\noindent\doublebox{\TheSbox}}

\definecolor{burntorange}{rgb}{0.8, 0.33, 0.0}
\definecolor{antiquefuchsia}{rgb}{0.57, 0.36, 0.51}
\definecolor{pomegranate}{RGB}{192, 57, 43}
\definecolor{light-gray}{gray}{0.95}

\lstnewenvironment{vuplisting}{%
  \lstset{%
    language=SSR,
    frame=none,
    literate=%
      {ρ}{\textrho}1%
      {Σ}{\textSigma}1%
      {Δ}{\textDelta}1%
      {η}{\texteta}1%
      {σ}{\textsigma}1%
      {<->}{{$\iff$}}1%
  }
}{}

\usepackage{amsmath,latexsym,amssymb}
\usepackage{amsfonts,amssymb}
\usepackage{mathtools}
\usepackage{algorithm}
\usepackage[noend]{algpseudocode}

\usepackage{paralist}

\usepackage{subfig}
\usepackage{multirow}
\usepackage[export]{adjustbox}

\usepackage{hyperref}
\newcommand{\urlNewWindow}[1]{\href[pdfnewwindow=true]{#1}{\nolinkurl{#1}}}

\usepackage{pgf}
\usepackage{tikz}
\usetikzlibrary{arrows,automata}
\newsavebox{\sembox}

\newtheorem{example}{Example}
\newtheorem{definition}{Definition}
\newtheorem{lemma}{Lemma}
\newtheorem{theorem}{Theorem}

\definecolor{grey}{rgb}{0.8,0.8,0.8}
\definecolor{darkgreen}{rgb}{0.4,0.7,0.4}
\definecolor{mydarkgreen}{rgb}{0.,0.4,0.1}
\definecolor{mydarkred}{rgb}{0.5,0,0}
\definecolor{darkblue}{rgb}{0.0, 0.0, 0.55}
\definecolor{tyrianpurple}{rgb}{0.4, 0.01, 0.24}
\definecolor{darkred}{rgb}{0.55, 0.0, 0.0}
\definecolor{mydarkgreen}{rgb}{0.0, 0.5, 0.0}
\definecolor{firebrick}{rgb}{0.69,0.13,0.13}

\usepackage[]{todonotes} 

\usepackage{xspace}

\newcommand{\C}{\lstinline[language=SSR,mathescape=true]}

\usepackage{xspace}
\newcommand{\coq}{Coq\xspace}

\newcommand{\ocaml}{OCaml\xspace}

\newcommand{\mathcomp}{Mathematical Components\xspace}

\newcommand{\bra}[1]{\ensuremath{\{#1\}}}

\newcommand{\rd}{RD\xspace}
\newcommand{\rdlong}{Regular Datalog\xspace}

\newcommand{\g}{\ensuremath{\mathcal{G}}}
\newcommand{\gdom}{\ensuremath{\mathbf{V}}}
\newcommand{\gsig}{\ensuremath{\mathbf{\Sigma}}}

\newcommand{\gedge}{\ensuremath{\mathbf{E}}}
\newcommand{\gvar}{\ensuremath{\mathcal{V}}}

\newcommand{\up}{\ensuremath{\Delta}}
\newcommand{\uP}{\ensuremath{\Delta_{+}}}
\newcommand{\uN}{\ensuremath{\Delta_{-}}}

\newcommand{\appd}[2]{\ensuremath{{#1}:\!\!+\!\!:{#2}}}
\newcommand{\upg}[3]{\ensuremath{{#1}\{{#2} \to {#3}\}}}
\newcommand{\upd}[4]{\ensuremath{{#1}\{{#2} \to ({#3},{#4})\}}}

\newcommand{\LD}[1]{\ensuremath{{#1}^{\Delta}}}
\newcommand{\LF}[1]{\ensuremath{{#1}^{\nu}}}

\newcommand{\MG}[2]{\ensuremath{{#1}^{\mathbf{#2}}}}
\newcommand{\MB}[1]{\MG{{#1}}{B}}
\newcommand{\MD}[1]{\MG{{#1}}{D}}
\newcommand{\MF}[1]{\MG{{#1}}{F}}

\newcommand{\lsat}[1]{\ensuremath{\g \models {#1}}}
\newcommand{\csat}[2]{\ensuremath{\g \models_{#1} {#2}}}

\newcommand{\dsat}[3][\up]{\ensuremath{\appd{\g}{{#1}} \models_{#2} {#3}}}

\newcommand{\supp}{\ensuremath{\mathsf{supp}}}
\newcommand{\sym}{\ensuremath{\mathsf{sym}}}

\newcommand{\sset}{\ensuremath{\Sigma}}

\newcommand{\iTp}{\ensuremath{T_{\g,\supp}^{\Pi}}}
\newcommand{\iTcl}[1]{\ensuremath{T_{\g,\supp}^{\Pi,#1}}}

\newcommand{\iTBcl}[1]{\ensuremath{T^{\Pi,#1}}}

\newcommand{\uset}[2]{\ensuremath{\{#1\}\cup{#2}}}

\newcommand{\matcha}[1][\g]{\ensuremath{M^A_{#1}}}
\newcommand{\matchl}{\ensuremath{M^L_{\g}}}
\newcommand{\matchb}[1][\g]{\ensuremath{M^B_{#1}}}

\newcommand{\matchda}{\ensuremath{M^{A,m}_{\g,\up}}}

\newcommand{\matchdb}{\ensuremath{M^{B}_{\g,\up}}}

\newcommand{\strl}{\ensuremath{\sset_\rhd}}
\newcommand{\strg}{\ensuremath{\sset_\lhd}}

\newcommand{\twd}{\ensuremath{\mathit{WD}}}
\newcommand{\tsnb}{\ensuremath{\mathit{SNB}}}

\newif\ifmore
\morefalse

\newcommand{\dg}{\ensuremath{\Delta\g}}

\usepackage{pifont}
\usepackage{marginnote}

\newcommand\fullinlinelink[2]{\href{#1}{#2}~\raisebox{-.2em}{\includegraphics[height=.9em]{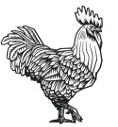}}}

\newcommand\baseNetAddress{https://github.com/VerDILog/VerDILog/tree/iclp-2018/html}
\newcommand\inlinelink[3][]{\fullinlinelink{\baseNetAddress/#2.html#1}{#3}}

\title[Certified Graph View Maintenance with \rdlong]{Certified Graph View Maintenance \\ with \rdlong}
\author[Bonifati and Dumbrava and Gallego Arias]{
  Angela Bonifati, Stefania Dumbrava \\
  LIRIS, Université Lyon 1, France
  \and
  Emilio Jesús Gallego Arias \\
  MINES ParisTech, PSL Research University, France
}

\begin{document}

\maketitle

\begin{abstract}
  We employ the Coq proof assistant to develop a
  mechanically-certified framework for evaluating graph queries and
  incrementally maintaining materialized graph instances, also called
  views. The language we use for defining queries and views is
  \rdlong(\rd) -- a notable fragment of non-recursive Datalog that can
  express complex navigational queries, with transitive closure as
  native operator. We first design and encode the theory of RD and
  then mechanize a RD-specific evaluation algorithm capable of
  fine-grained, incremental graph view computation, which we prove
  sound with respect to the declarative RD semantics.
  By using the Coq extraction mechanism, we test an \ocaml version of
  the verified engine on a set of preliminary benchmarks.
  Our development is particularly focused on leveraging existing
  verification and notational techniques to: a) define mechanized
  properties that can be easily understood by logicians and database
  researchers and b) attain formal verification with limited
  effort. Our work is the first step towards a unified,
  machine-verified, formal framework for dynamic graph query languages
  and their evaluation engines.

\end{abstract}

\section{Introduction}

\label{sec:introduction}
Modern graph query engines\footnote{Several successful commercial
implementations exist, e.g., Neo4j~\cite{Neo4j}, Google's Cayley~\cite{Cayley}, 
Twitter's FlockDB~\cite{FlockDB} and Facebook's Apache Giraph~\cite{Giraph}.} are gaining momentum, due to the proliferation of \emph{interconnected} data 
and related applications, spanning from social networks to scientific databases and the Semantic Web. 
The adopted query languages are \emph{navigational}, focusing on data topology and, particularly, on \emph{label-constrained reachability}. 
Notable examples~\cite{AnglesABHRV17} include Neo4j's
openCypher~\cite{Cypher}, Facebook's GraphQL~\cite{GraphQL}, SPARQL~\cite{SPARQL}, 
Gremlin~\cite{Gremlin}, Oracle's PGX~\cite{oraclepgx}, and the recent
G-CORE~\cite{gcore}, even though a standard graph query language is 
still undefined. 
At a foundational level, these languages are based on \emph{conjunctive
queries} (CQ), corresponding to Datalog~\cite{Ceri} clauses, i.e., function-free Horn formulas.
Their common denominator is that they perform \emph{edge traversals} (through join chains), while specifying and testing for the existence of label-constrained 
paths. 
Recently, a solution to the long standing open problem of identifying 
a suitable graph query Datalog fragment, balancing expressivity
and tractability, has been proposed in~\cite{Reutter2017}. We  
call this \emph{Regular Datalog} (\rd), a binary linear Datalog subclass that allows for complex, regular expression patterns between nodes.
\rd provides additional properties over full Datalog. First, its evaluation has NLOGSPACE-complete complexity, belonging to 
the NC class of \emph{highly parallelizable} problems. Second, RD query containment is \emph{decidable}, with an elementary tight 
bound (2EXPSPACE-complete)~\cite{Reutter2017}.

To the best of our knowledge, no \emph{specific evaluation algorithm}
for \rd queries has been proposed yet. What are the main desiderata
that such an algorithm should have?
On the one hand, real-world graphs are highly dynamic, ever-increasing in
size, and ever-changing in content. Hence, \emph{efficiency}, i.e., accelerating graph-based
query processing, has become more relevant than ever. On the other hand,
the sensitive nature of the data that large-scale, commercial graph
database engines operate on makes ensuring \emph{strong reliability
  guarantees} paramount.

We argue that having a deeply-specified framework for performing
\emph{incremental graph computation}, based on logic programming,
opens important perspectives in a variety of areas, especially in
light of security-sensitive applications involving graph-shaped
topologies, e.g., financial transaction, forensic analysis, and
network communication protocols.

\paragraph{Problem Statement:}
We target both requisites by developing a \textbf{mechanically-verified} framework for the
\emph{fine-grained} \textbf{incremental view maintenance of graph databases (IVMGD)}.
Specifically, let $\g$ be a graph database instance, $\dg$, a set of
modifications, and $V[\g]$, the materialization of an \rd view
\emph{or query} over $\g$. We provide an IVMGD-aware engine that
computes an \emph{incremental view update} $\Delta V[\g; \dg]$, making
the resulting view consistent with the updated graph database,
i.e., $V[\g] \cup \Delta V[\g; \dg] = V[\g \cup \dg]$.


\paragraph{Contributions:}
We build our mechanically-certified engine using \emph{theorem
  proving} techniques, in particular we use the \coq proof
assistant~\cite{coqref} to develop both the theory of the \rd language
and the evaluation engine itself. We make three major contributions:
a) we \emph{formalize}, in \coq, the syntax and semantics of Regular
Queries, as Regular Datalog programs; b) we \emph{implement}, in
\coq's functional programming language Gallina, an executable engine
for incremental \rd maintenance; c) we \emph{prove} the engine is
\emph{sound}, i.e. that it correctly computes incremental view
updates.

We encode the semantics of \rd using the finite set theory in the
\mathcomp library, which was developed to carry out the mechanized
proof of the Feit-Thompson theorem~\cite{GonthierFT} on finite group
classification; it thus provides excellent support for finite
reasoning.
This brings the actual written-down mechanized semantics very close to
a mathematical language, making it more accessible to non-expert
practitioners -- a central point to understand the guarantees provided
by our development.

To develop our incremental graph view maintenance algorithm, we
adapt the classical delta-rule one for IVM~\cite{Gupta1993} --- initially
designed for non-recursive SQL --- to the \rd evaluation setting.

Lastly, we prove our main result: ``Let $\g$ be a base instance and
$\Pi$, a \rd program of view $V$. If $\Pi$ is satisfied by the
materialized view $V[\g]$, then, for a $\g$ update $\dg$, the IVMGD
engine outputs an \emph{incremental view update}, $\Delta V[\g; \dg]$,
such that $\Pi$ is satisfied by $V[\g] \cup \Delta V[\g; \dg]$''. The
proof relies on two key, newly-developed mechanized theories for
\emph{stratified} and \emph{incremental} satisfaction of Datalog
programs.

As mentioned in~\cite{FanHT17}, theoretical work on graph view
maintenance is still in its infancy and, as noted in~\cite{BeyhlG16},
most mainstream commercial graph databases do not provide concepts for
defining graph views and maintenance.\footnote{The only recent
  exception -- which can handle named queries and updates -- is Cypher
  for Apache Spark
  \url{https://github.com/opencypher/cypher-for-apache-spark}.}
Thus, we believe that our verified engine builds the foundations for
certifying realistic graph query and update engines, using declarative
paradigms, such as Datalog subsets, as their query language. Specifically,
this certified specification could serve as a blueprint for future graph 
query design and ongoing standardisation
efforts~\cite{gcore}. Additionally, we consider that most of our
verification techniques are not restricted to the \rdlong setting, but
are also applicable to broader logic programming contexts.

\paragraph{Organization:}
The paper is organized as follows. In Section~\ref{sec:rq}, we
illustrate the syntax and semantics of \rd and its \coq
formalization. In Section~\ref{sec:engine}, we present the IVMGD
algorithm and, in Section~\ref{sec:proof}, we summarize its mechanized
proof. Section~\ref{sec:eval} shows our extracted engine's performance
on graph datasets with synthetic queries in RD. We describe related
work in Section~\ref{sec:rel} and conclude and outline perspectives in
Section~\ref{sec:concl}. The \coq code for this paper
can be downloaded from: \url{https://github.com/VerDILog/VerDILog/tree/iclp-2018}.


\section{\rdlong: Design and Formalization}\label{sec:rd}
\label{sec:rq}
%
%
In this section, we present the theory of \emph{Regular Datalog} (RD) and its \coq mechanization. The language is based on Regular Queries (RQs)~\cite{Reutter2017}. 
In Sec.~\ref{sec:rqsyn} and Sec.~\ref{sec:rqsem}, we detail our encoding of RD sytax and semantics. 
Sec.~\ref{sec:rqex} illustrates potential usages of the language, in the context of financial transaction and social network graphs.
\subsection{\rd Syntax}\label{sec:rqsyn}
We fix the finite sets of constants (nodes) and symbols (edge labels), namely, $\mathbf{V}$ and $\Sigma$.
\begin{definition}[Graph Database]
  A \emph{graph instance} $\g$ over $\gsig$ is a set of \emph{directed} labelled edges, $\gedge$, where
  $\gedge \subseteq \gdom \times \gsig \times \gdom$.
  A \emph{path} $\rho$ of length $k$ in $\g$ is a sequence $n_1 \xrightarrow{s_1} n_2 \ldots n_{k-1} \xrightarrow {s_k}
  n_k$. Its label is the concatenation of edge symbols, i.e., $\lambda(\rho) = s_{1} \ldots s_{k} \in \gsig^{*}$.
  %
\end{definition}
$\g$ can be seen as a database
$\mathcal{D}(\g) = \{ s(n_1,n_2) \mid (n_1, s, n_2) \in \gedge \}$,
by interpreting its edges as binary \emph{relations} between nodes. 
$\mathcal{D}(\g)$ is also called the \emph{extensional database} (EDB).

In order to model graphs in \coq, we assume a pair of \emph{finite types}
(\C|finType|), representing edge labels and nodes. The graph encoding (\inlinelink[\#egraph]{VUP.vup}{\C|egraph|})
is, thus, a \emph{finitely supported function} (\inlinelink[\#lrel]{VUP.vup}{\C|lrel|}), mapping
labels to their corresponding set of edges:
%
%
%
\begin{vuplisting}
Variables (V Σ : finType).
Inductive L := Single | Plus.
Inductive egraph := EGraph of {set V * V}.
Inductive lrel   := LRel of {ffun Σ * L -> egraph}
\end{vuplisting}
Note that, for each label, our graph representation maintains both the
regular set of edges and its transitive closure, denoted by the \C|L|
type. We explain its usage in later sections.
%
%
%
\begin{definition}[\rdlong (\rd)]
  \emph{Regular Datalog} is the binary --- all atoms have arity 2 ---
  Datalog fragment, with \emph{recursion restricted to transitive
    closure} and \emph{internalized} as labels on literals.
\end{definition}
There are several approaches to making the representation of logic
programs amenable to efficient mechanical reasoning. We have found
that indexing \emph{completed} clauses by \emph{head} symbols works
well, as the corresponding canonical disjunctive definitions, from
the clause map, become readily accessible in proofs.
Fig.~\ref{fig:rq} provides the formal syntax for \rdlong programs. A
program is a map from each symbol in $\Sigma$, to a \emph{single} clause head and
\emph{normalized} disjunctive body. The normalized form is obtained by first
transforming all clauses to a common head representation and then 
grouping their respective bodies. This classical process is similar to the completion
procedure in~\cite{adbt/Clark77}. For example, the program:
%
%
$s(a,b).\; s(z,y) \leftarrow p(x,y), q^+(z,x)$ is normalized as
$s(x,y) \leftarrow (a = x \land b = y) \lor (p(z,y) \land q^+(x,z))$
and represented by a \emph{function} from $s$ to the head and
disjunctive body.
%
%
\begin{figure}
  \begin{equation*}
  \begin{array}{r@{\quad\color{red}::=\quad}l@{\qquad\qquad}r}
    t & n \in \gdom \mid x \in \gvar  & \text{(Terms, Node ids)}   \\
    A & s(t_1, t_2), \text{ where } s \in \Sigma \mid t_1 = t_2  & \text{(Atoms)},  \\
    L & A \mid A^+                               & \text{(Literals)}             \\
    B & L_1 \land \ldots \land L_n               & \text{(Conjunctive Body)}     \\
    D & B_1 \lor \ldots \lor B_n                 & \text{(Disjunctive Body)}     \\
    C & (t_1, t_2) \leftarrow  D                     & \text{(Clause)}              \\
    \Pi & \Sigma \to \{C_1, \ldots, C_n \}           & \text{(Program)} \\
  \end{array}
  \end{equation*}
  \caption{Regular Datalog Grammar}
  \label{fig:rq}
\end{figure}
We encode \rd primitives in \coq as records:
\begin{vuplisting}
Record atom    := Atom    { syma : Σ; arga : T * T }.
Record lit     := Lit     { tagl : L; atoml: atom }.
Record cbody   := CBody   { litb : seq lit }.
Record clause  := Clause  { headc: T * T; bodyc : seq cbody }.
Inductive program := Program of {ffun Σ -> clause T Σ L}.
\end{vuplisting}
A key feature of this formalization is that it is parametric in the
variables \C|$\Sigma$,T,L|. This design choice allows sharing the representation
for ground and non-ground clauses, and is central to the incrementality
proof, in which we will decorate literals with customs labels. Also, note our
naming convention for \coq constants, whereby the last letter denotes
the type. For example, \inlinelink[\#syma]{VUP.vup}{\C|syma|} is the function that
returns the symbol for an atom. \inlinelink[\#syml]{VUP.vup}{\C|syml|} does the same for a literal. Similarly,
satisfaction conditions will be named \C|sTa|, \C|sTl|, etc., depending on
the argument type.

\begin{definition}[Regular Queries (RQ)]
  A \emph{regular query} $\Omega$ over $\g$ is a \rd program $\Pi$,
  together with a distinguished query clause, whose head is the
  top-level \emph{view} ($V$) and whose body is a conjunction of $\Pi$
  literals.
\end{definition}

\subsection{\rd Semantics}\label{sec:rqsem}
%
%
The semantics of \rd programs follows a standard term-model
definition. As noted in Sec.~\ref{sec:rqsyn}, for optimization
purposes, interpretations $\g$ are modeled as \emph{indexed relations}
$(\gsig \times \bra{\emptyset,+}) \to \mathcal{P}(\gdom \times \gdom)$, containing labeled graphs
and their transitive closure.
%
Then, program satisfaction builds on the below definition of
satisfaction for (ground) literals:
\begin{definition}[Literal Satisfaction]\label{def:sem:int-literal}
  The satisfaction $\lsat{L}$ of a ground literal $L = s^l(n_1,n_2)$ is defined as:
\begin{equation*}
  \begin{array}{lcl}
    \lsat{s^l(n_1,n_2)}   & \iff & (n_1,n_2) \in \g(s,l) \\
  \end{array}
\end{equation*}
Note that, in order for this definition to be correct, $\g$ must be
\emph{well-formed}, that is to say, the information stored in
$\g(s,+)$ has to correspond to the actual transitive
closure of $\g(s,\emptyset)$. We can state this condition as:
\begin{equation*}
  \begin{array}{lcl}
    \mathsf{wfG}(\g) &\iff& \forall s,  \mathsf{is\_closure}(\g(s,\mathsf{\emptyset}),\g(s,+)) \\
    \mathsf{is\_closure}(g_s,g_p) &\iff& \forall (n_1,n_2) \in g_p,
    \exists \rho \in \gdom^{+}, \mathsf{path}(g_s,n_1,\rho) \land \mathsf{last}(\rho) = n_2 \\
    \mathsf{path}(g,n_1,\rho) &\iff& \forall i \in \{1\ldots|\rho|\}, (n_i, n_{i+1}) \in g
  \end{array}
\end{equation*}
where the node list $\rho \in \gdom^+$ represents the path without
including the initial node $n_1$. 
\end{definition}
%
%
%
Note that we compile the surface syntax $s^{-}(X,Y)$ to $s(Y,X)$ and $s^*(X,Y)$ to
$X=Y \lor s^+(X,Y)$. The \coq encoding \inlinelink[\#sTl]{VUP.vup}{\C|sTl gl|} of $\lsat{L}$ is
a direct transcription of Def.~\ref{def:sem:int-literal}, where we henceforth
omit the \C|G| parameter, as is it assumed to be implicit:
\begin{vuplisting}
Definition sTl gl := G (syml gl, tagl gl) (argl gl).1 (argl gl).2.
\end{vuplisting}
\begin{definition}[Clause Satisfaction]\label{def:sem:int-clause}
  A \rd clause with disjunctive body
  $D \equiv (L_{1,1} \land \ldots \land L_{1,n}) \lor \ldots \lor
  (L_{m,1} \land \ldots \land L_{m,n})$ and head symbol $s$ is
  satisfied by $\mathcal{G}$ iff, for all groundings $\eta$, whenever the
  corresponding instantiation of a body in $D$ is satisfied, then the head is also satisfied. Formally:
  \begin{equation*}
    \begin{array}{rc}
      \csat{s}{(t_1,t_2) \leftarrow (L_{1,1} \land \ldots \land L_{1,n}) \lor \ldots \lor
        (L_{m,1} \land \ldots \land L_{m,n})} & \iff \\
      \forall \eta,
      \bigvee_{i=1..m} (\bigwedge_{j=1..n} \lsat{\eta(L_{i,j}))} \Rightarrow \eta(s(t_1,t_2))
    \end{array}
  \end{equation*}
\end{definition}
The \coq encoding of Def.~\ref{def:sem:int-clause} relies on the
definition of an instantiation with a \emph{grounding} $\eta$. We model 
$\eta$ as a function \C|g|, of type \C|gr|, mapping from $\gvar$ (the ordinal type \C|'I_n|) to $\gdom$.
This extends straightforwardly from terms to clauses.
\begin{vuplisting}
Definition gr n := {ffun 'I_n -> V}.
Definition sTb b := all sTl (litb b).
Definition sTc n s c := [forall g : gr n, let gc := grc g c in
			 has sTb (bodyc gc) ==> G (s,Single) (headc gc)].
\end{vuplisting}
Note that \C|has| and \C|all| are the counterparts of the corresponding logical operations, extended
to lists.
We can define a model for an \rd program as:
\begin{definition}[Program Satisfaction]\label{def:sem:int-program}
  A \emph{well-formed} interpretation $\g$ is a model for a program
  $\Pi$ with respect to $\Sigma$ iff $\g$ satisfies all of the $\Pi$
  clauses indexed by symbols in $\Sigma$:
  $\g \models_{\Sigma} \Pi \iff \forall~s \in \Sigma,
  \csat{s}{\Pi(s)}$.
\end{definition}
The \coq encoding is straightforward:
\begin{vuplisting}
Definition sTp (p : program n) := [forall s, sTc n s (p s)].
\end{vuplisting}
%
The formalized definition of satisfaction in
Def.~\ref{def:sem:int-program} is crucial to understanding the main
soundness theorem in Sec.~\ref{sec:sound}. It establishes that the
output of the mechanized engine is a \emph{model} of the input
program, i.e., that it complies to the satisfaction specification of
Def.~\ref{def:sem:int-program} above. Hence, if this definition were to be
incorrect --- for example, by making \C|sTp p = true| -- the
theorem would become meaningless.
%
%
\subsection{\rd Examples}\label{sec:rqex}
We write $(r+s)(x, y)$ for $r(x,y) \lor s(x,y)$ and
$(r \cdot s)(x,y)$ for $r(x,z) \land s(z,y)$.
\begin{figure}
  \centering
  \subfloat[Potential Fraud]{
    \begin{tikzpicture}[scale=3, every node/.style={scale=0.7}]
      \tikzstyle{every state}=[fill=none,draw=black,text=black, minimum size=0.5mm, inner sep=1pt]
      \node[state]         (A) at (0,0)      {$X$};
      \node[state]         (B) at (0.75,0)   {$Y$}; 
      \path (A)  edge[->, bend left = 15, sloped, above]  node {\small{$\mathbf{pstransfer}^{+}$}}  (B)
            (B)  edge[->, bend left = 15, sloped, below]  node {\small{$\mathbf{pstransfer}^{+}$}}  (A);
     \end{tikzpicture}
  }
  \qquad
  \subfloat[Secured Transfer]{
    \begin{tikzpicture}[scale=1.45, every node/.style={scale=0.7}]
  \tikzstyle{every state}=[fill=none,draw=black,text=black, minimum size=0.5mm, inner sep=0.8pt]
  \node[state]         (B1) at (-0.5,0)      {$X$};
  \node[state]         (B2) at (0,0)       {};
  \node[state]         (B3) at (0.5,0)       {};
  \node[state]         (B4) at (1,0)       {};
  \node[state]         (B5) at (1.5,0)       {};
  \node[state]         (B6) at (2,0)       {$Y$};
  \node[state]         (V)  at (0.75,0.75) {$Z$};
  \path (B1) edge[->,above right, color=blue]  node {\small{\textbf{c}}}  (B2)
        (B2) edge[->,above, color=blue]  node {\small{\textbf{c}}}  (B3)
        (B3) edge[->,dashed,color=blue]  node {}   (B4)
        (B4) edge[->,above, color=blue]  node {\small{\textbf{c}}}  (B5)
        (B5) edge[->,above left, color=blue]  node {\small{\textbf{c}}}  (B6)
        (B2) edge[<-, sloped,left, above, color=blue]  node {\small{$\mathbf{m}^{+}$}}  (V)        
        (B3) edge[<-, sloped,left, above, color=blue] node {\small{$\mathbf{m}^{+}$}}  (V)
        (B4) edge[<-,sloped,right,above, color=blue] node {\small{$\mathbf{m}^{+}$}}  (V)
        (B5) edge[<-, sloped,right,above, color=blue] node {\small{$\mathbf{m}^{+}$}}  (V)
        (B6) edge[->,bend left=6, below] node {\small{\textbf{a}}}  (B1)  
        (V) edge[->,bend right=10, above, color=blue] node {\small{\textbf{a}}}  (B1)  
        (B1) edge[->,bend left=20, below] node {\small{$\mathbf{t}$}}  (B6);
  \end{tikzpicture}
 }
 \caption{Fraud Detection}
 \label{fig:fraud-detection}
\end{figure}
\begin{example}[Fraud Detection]\label{ex:fd}
Consider a financial transaction network, in which entities can \textbf{c}onnect and \textbf{t}ransfer money to each other, as well as \textbf{m}onitor and 
\textbf{a}ccredit each other. A potential fraud/\emph{suspect transaction}, e.g., money laundering, is a cycle of $\mathbf{pstransfer}^{+}$, i.e., potentially secured transfer 
chains (Fig.~\ref{fig:fraud-detection}). A potentially secured transfer ($\mathbf{pstransfer}$) is either a transfer ($\mathbf{t}$) or a \emph{secured transfer}.
A \emph{secured transfer} from $X$ to $Y$ occurs if $Y$ is accredited ($\mathbf{a}$) by $X$ and if $X$ {\color{blue}secures} a connection and transfers ($\mathbf{t}$) to $Y$. 
$X$ {\color{blue}secures} a connection to $Y$ (blue subgraph in Fig.~\ref{fig:fraud-detection}), if it connects (${\color{blue}\mathbf{c}}$) via a chain of intermediaries that are centrally monitored 
(${\color{blue}\mathbf{m^{+}}}$) by an accredited (${\color{blue}\mathbf{a}}$) entity $Z$. Potentially fraudulent (\emph{suspect}) transactions can be computed with the \rd program:
{\small
\[
\begin{array}{l@{}l@{~}l}
\mathbf{s}uspect(X,Y) & \leftarrow & \mathbf{pstransfer}^{+}(X,Y), \mathbf{pstransfer}^{+}(Y,X) \\
\mathbf{pstransfer}(X,Y) & \leftarrow & (\mathit{\mathbf{t}ransfer} + \mathit{stransfer})(X,Y) \\
stransfer(X,Y) & \leftarrow & \mathit{\mathbf{a}ccredited} (Y,X), \mathit{secures}(X,Y), \mathit{\mathbf{t}ransfers}(X,Y) \\
secures(X,Y) & \leftarrow & (\mathit{\mathbf{c}onnected} \cdot \mathit{cmonitored}^{+} \cdot \mathbf{c}onnected)(X,Y) \\
cmonitored(X,Y) & \leftarrow & \mathit{\mathbf{c}onnected}(X,Y), \mathit{\mathbf{m}onitors}^{+}(Z,X), \mathit{\mathbf{m}onitors}^{+}(Z,Y), \mathit{\mathbf{a}ccredited}(Z,X)\\
\end{array}
\]
}
\end{example}

\begin{example}[Brand Reach]
Consider a platform, such as Twitter, with asymmetric connections and, thus, an underlying directed graph topology.
Let $Z$ be a brand (central node in Fig.~\ref{fig:br}) that wants to determine its \emph{\textbf{r}each}, i.e. 
\emph{potential clients} pairs (empty nodes in Fig.~\ref{fig:br}). 
A pair of users ($X$, $Y$) are \emph{potential clients}, if both were \emph{exposed} to $Z$. 
We say $X$ is \emph{exposed} to $Z$, if $X$ \emph{\textbf{e}ndorses} $Z$ or if it is connected, through a 
potential chain of followers, to an influencer that \emph{\textbf{e}ndorses} $Z$. We say a 
user \emph{\textbf{e}ndorses} a brand, if it \emph{likes} and \emph{advertises} the brand.
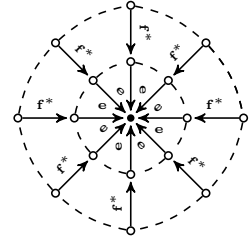
\begin{figure}[h!]
  \captionsetup[subfloat]{farskip=-14pt,captionskip=1pt}
  \adjustbox{valign=t}{
  \subfloat{  
  \small{
    $
    \begin{array}{lll}
      \mathbf{r}each(X,Y) & \leftarrow & (pclients + pclients^{-})(X,Y) \\
      pclients(X,Y)       & \leftarrow & exposed(X,Z), exposed(Y,Z) \\
      exposed(X,Z)        & \leftarrow &(\mathbf{f}ollows^{*} \cdot \mathbf{e}ndorses)(X,Z) \\
      \mathbf{e}ndorses(X,Z) & \leftarrow & likes(X,Z), advertises^{+}(X,Z)
    \end{array}
    $}
  }}
  \qquad 
  \adjustbox{valign=t}{
  \subfloat{\small   
  \begin{tikzpicture}[scale=0.5,->,>=stealth',shorten >=1pt,node distance=2cm,semithick]
  \tikzstyle{every state}=[fill=black,draw=none,text=white, minimum size=0.5mm, inner sep=1pt]
  \node[state]         (B)  at (0,0)       {}; 
  \node[state,draw,fill=white] (I1) at (0,1.5)     {};
  \node[state,draw,fill=white] (I2) at (1.5,0)     {};
  \node[state,draw,fill=white] (I3) at (0,-1.5)    {};
  \node[state,draw,fill=white] (I4) at (-1.5,0)    {};
  \node[state,draw,fill=white] (F1) at (0,3)       {};
  \node[state,draw,fill=white] (F2) at (3,0)       {};
  \node[state,draw,fill=white] (F3) at (0,-3)      {};
  \node[state,draw,fill=white] (F4) at (-3,0)      {};
  \node[state,draw,fill=white] (F1') at (2,2)      {};
  \node[state,draw,fill=white] (F2') at (2,-2)     {};
  \node[state,draw,fill=white] (F3') at (-2,-2)    {};
  \node[state,draw,fill=white] (F4') at (-2,2)     {};
  \node[state,draw,fill=white] (I1') at (1,1)   {};
  \node[state,draw,fill=white] (I2') at (1,-1)  {};
  \node[state,draw,fill=white] (I3') at (-1,-1) {};
  \node[state,draw,fill=white] (I4') at (-1,1)  {};
  \path (I1) edge[sloped,above]        node {\tiny{$\mathbf{e}$}} (B)
	(I2) edge[sloped,below]        node {\tiny{$\mathbf{e}$}} (B)
	(I3) edge[sloped,above]        node {\tiny{$\mathbf{e}$}} (B)
	(I4) edge[sloped,above]        node {\tiny{$\mathbf{e}$}} (B)
        (I1') edge[sloped,below]        node {\tiny{$\mathbf{e}$}} (B)
	(I2') edge[sloped,below]        node {\tiny{$\mathbf{e}$}} (B)
	(I3') edge[sloped,above]        node {\tiny{$\mathbf{e}$}} (B)
	(I4') edge[sloped,above]        node {\tiny{$\mathbf{e}$}} (B)
	(F1) edge[sloped,above]        node {\tiny{$\mathbf{f}^{*}$}} (I1)
	(F2) edge[sloped,above]        node {\tiny{$\mathbf{f}^{*}$}} (I2)
	(F3) edge[sloped,above]        node {\tiny{$\mathbf{f}^{*}$}} (I3)
	(F4) edge[sloped,above]        node {\tiny{$\mathbf{f}^{*}$}} (I4)
	(F1') edge[sloped,above]        node {\tiny{$\mathbf{f}^{*}$}} (I1')
	(F2') edge[sloped,above]        node {\tiny{$\mathbf{f}^{*}$}} (I2')
	(F3') edge[sloped,above]        node {\tiny{$\mathbf{f}^{*}$}} (I3')
	(F4') edge[sloped,above]        node {\tiny{$\mathbf{f}^{*}$}} (I4')
	(F1)  edge[-, bend left=15, dashed, sloped, above] node {} (F1') 
	(F1') edge[-, bend left=15, dashed, sloped, above] node {} (F2) 
	(F1') edge[-, bend left=15, dashed, sloped, above] node {} (F2) 
	(F2)  edge[-, bend left=15, dashed, sloped, above] node {} (F2') 
	(F2') edge[-, bend left=15, dashed, sloped, above] node {} (F3) 
	(F3)  edge[-, bend left=15, dashed, sloped, above] node {} (F3') 
	(F3') edge[-, bend left=15, dashed, sloped, above] node {} (F4) 
	(F4)  edge[-, bend left=15, dashed, sloped, above] node {} (F4') 
	(F4') edge[-, bend left=15, dashed, sloped, above] node {} (F1) 
	(I1) edge[-, bend left=15, dashed, sloped, above] node {} (I1') 
	(I1') edge[-, bend left=15, dashed, sloped, above] node {} (I2) 
	(I2)  edge[-, bend left=15, dashed, sloped, above] node {} (I2') 
	(I2') edge[-, bend left=15, dashed, sloped, above] node {} (I3) 
	(I3)  edge[-, bend left=15, dashed, sloped, above] node {} (I3') 
	(I3') edge[-, bend left=15, dashed, sloped, above] node {} (I4) 
	(I4)  edge[-, bend left=15, dashed, sloped, above] node {} (I4') 
	(I4') edge[-, bend left=15, dashed, sloped, above] node {} (I1) 
	;
   \end{tikzpicture}
 }}
 \caption{Brand Reach}
 \label{fig:br}
\end{figure}

\end{example}

\section{\rdlong Evaluation: A Mechanized IVMGD-aware Engine}\label{sec:engine}
%
We now describe the design of the mechanized IVMGD engine, which is
based on the non-recursive bottom-up evaluation of \rd programs. We first
describe the top-level interface of the engine and the main execution
loop. Then, in Sec.~\ref{sec:engine:base}, we outline the building
block of \emph{non-incremental} clause evaluation; in
Sec.~\ref{sec:engine:delta-join}, we describe the delta-join
\emph{incremental clause evaluation} algorithm. Finally,
Sec.~\ref{sec:engine:delta} explains the implementation of the
delta-join heuristic in the actual engine.

\subsection{Top-level Interface and Overview}
\label{sec:engine:toplevel}
The incremental \rd evaluation engine is designed around bottom-up
non-recursive program model computation. This is indeed adequate for
graph and regular queries as these internalize recursion using
\emph{closure operations} --- usually computed by specialized tools
optimized for efficiency. While our engine builds on ideas
from~\cite{BCD17}, we have considerably redesigned all its core
components: syntax is now based on a new \emph{parametric} and
\emph{normalized} representation, and the core evaluation
infrastructure and theory have been redesigned to account for
\emph{stratified}, \emph{single-pass}, non-recursive
\emph{incremental} model computation.  The formalization workload is
higher in our setting, as we cannot rely on the usual fixpoint
theorems, but must define a custom theory for modular program
satisfaction.

A key problem to solve when reasoning about \emph{incremental}
computation is the representation of changes. Indeed, a formal
definition of updates can be delicate to state, as it must
account for potentially overlapping additions and removals,
order issues, etc. To this end, we define
\emph{canonical} graph updates, reminiscent of, but weaker than
``change structures'' --- defined in~\cite{Cai:2014:TCH:2594291.2594304}:
\begin{definition}[Graph Updates]
  An \emph{update} $\up \equiv (\uP, \uN)$ is a pair of
  \emph{disjoint} graphs, respectively representing insertions and
  deletions.
  %
\end{definition}
In \coq, updates are encoded as a record \inlinelink[\#edelta]{VUP.vup}{\C|edelta|} packing the graphs and a
disjointedness proof; this allows us to consider only well-formed
updates. Note that \C|addd| and \C|deld| are the record fields representing
$\uP$ and $\uN$:
\begin{vuplisting}
Definition wf_edelta addd deld := [forall s, [disjoint addd s & deld s]].
Structure edelta := { addd : lrel; deld : lrel; _ : wf_edelta addd deld }.
\end{vuplisting}
The core operations for updates are their \emph{application} to a base graph, and
the \emph{modification} of the changes pertaining to a particular symbol:
\begin{definition}[Update application and modification]\label{def:gmod}
  \begin{equation*}
    \begin{array}{lclr}
      \appd{\g}{\up}        &\equiv& \g \setminus \uN \cup \uP  & (\textit{application})\\
      \upd{\up}{s}{g_+}{g_-} &\equiv& (\upg{\uP}{s}{g_+},
      \upg{\uN}{s}{g_{-} \setminus g_+}) & (\textit{modification})
    \end{array}
  \end{equation*}
\end{definition}
Armed with these definitions, plus those corresponding to the \rd syntax
and semantics from Sec.~\ref{sec:rd}, we can now define the top-level
interface to our engine.
A particular challenge that arose during the development of the
interface was allowing repeated, incremental $\up$-aware
calls. Achieving composition proved to be quite challenging, as
the soundness invariant must be preserved along calls. In total, six
parameters had to be used.
The \emph{static} input parameters are: a program $\Pi$, a graph $\g$,
and a set of symbols, or \emph{support} $\supp$, which indicates the
validity of a subset of $\g$, and thus what information the
incremental engine may not recompute. Indeed, a precondition of the
engine is that the input graph must be a model of $\Pi$ up to $\supp$,
that is to say, $\csat{\supp}{\Pi}$.
Note that in the database literature, $\gsig$ is usually seen as a
disjoint set pair, ($\gsig_E,\gsig_I$), corresponding to the \emph{extensional}
and \emph{intensional} parts of a program. For our engine, this
distinction is ``dynamic'', in the sense that an already-processed
strata-level is seen as ``extensional'', or immutable, for the rest of
the execution.
Thus, typical cases for $\supp$ will be $\supp \equiv \gsig_E$, when
the engine has been never been run before, or $\supp \equiv \gsig$,
where $\g$ is the output of a previous run, and thus the consequences
for all clauses have been computed.
%

Whereas the program, graph, and support are fixed during the execution of
the engine, the latter will take an additional three \emph{dynamic} parameters,
representing the current execution state. These are: $\up$, the current update,
which is modified at each iteration, and $\strl, \strg$, which respectively
represent the set of processed symbols/stratum and the ``todo'' list.
We write $\up_O = \iTp(\strl,\strg,\up)$ (written \inlinelink[\#fwd_program]{VUP.vup}{\C|fwd_program|}
in \coq) for a call to the engine returning an update $\up_O$. We
prove that the engine implements a \emph{program consequence
  algorithm}, thus satisfying $\dsat[\up_O]{\gsig}{\Pi}$.
Assuming a \emph{clause consequence operator} $\iTcl{s}(\up)$ (or
\inlinelink[\#fwd_or_clause]{VUP.vup}{\C|fwd_or_clause|}), the below
\coq implementation of the engine iterates over the unprocessed symbol
list $\strg$ and, for each of its symbols \C|s| to be inspected, it
computes a new update $\up'$. To this end, it modifies $\up$ with the
consequences for the clause indexed by \C|s| and computes the
corresponding closure of \C|s|. The algorithm then makes the recursive
call, adding \C|s| and \C|s+| to the set of processed symbols.
\begin{vuplisting}
Fixpoint fwd_program $\Pi$ G supp Δ $\strl$ $\strg$ : edelta := match $\strg$ with
 | [::]        => Δ
 | [:: s & ss] =>
   let (arg, body) := $\Pi$ s                       $\!$in
   let Δ'  := fwd_or_clause G supp Δ s arg body in
   let Δ'  := compute_closures G Δ' s           in
   fwd_program $\Pi$ G supp Δ' (s $\cup$ s+ $\cup$ $\strl$) ss
\end{vuplisting}
Note that the \C|compute_closures| function above is an abstraction
over an arbitrary algorithm for closure computation, that we assume
correct. For instance, we hope to use the verified implementation of
Tarjan's algorithm from~\cite{Tarjan}.
%

As can be observed from the above code, the core part of the algorithm
is concentrated in clause evaluation. Hence, we now proceed to
presenting \emph{base} and \emph{incremental} for clause evaluation.
Note that the base --- or \emph{non-incremental} --- method is still
needed since in some cases, incremental evaluation is either not
possible or not sensible, as full re-computation may be faster.

\subsection{The Base Engine for \rd Evaluation}
\label{sec:engine:base}

The \emph{base}, or \emph{non-incremental}, clause evaluation
implements a forward-chaining \emph{consequence operator}, using a
\emph{matching algorithm} $M$; this takes as input atoms, literals, or
clause bodies and returns a set of substitutions, as explained
in~\cite{Alice}.  Basic literal matching, $\matchl(l)$
(\inlinelink[\#match_lit_all]{VUP.vup}{\C|match_lit|} in \coq) takes a literal
and returns the set of all substitutions $ss$ that satisfy it, so that
$\forall \sigma \in ss,~\lsat{\sigma(l)}$. Literal matching is
extended to body matching in an straightforward way, with $\matchb(B)$
(\inlinelink[\#match_body]{VUP.vup}{\C|match_body|}) traversing $B$ and
accumulating the set of substitutions obtained from the individual
matching.
The algorithm we consider corresponds to computing \emph{nested-loop} join
and we implement it in a functional style, using a monadic fold.
Substitutions are then accumulated for each disjunctive clause and
\emph{grounded} heads are added to the interpretation:
\begin{definition}[Clausal Consequence Operator]
  \label{def:tpc}
  Given a \rd clause
  $\Pi(s) \equiv (t_1, t_2) \leftarrow \bigvee_{i=1..n} B_i$, the
  \emph{base clausal consequence operator} $\iTBcl{s}(\g)$ computes the set of
  facts that can be \emph{inferred} from $\g$:
  \begin{equation*}
    \iTBcl{s}(\g) \equiv \bra{\sigma(t_1, t_2) \mid \sigma
    \in \bigcup_{i=1..n} \matchb(B_i) }.
  \end{equation*}
\end{definition}
The \inlinelink[\#fwd_or_clause]{VUP.vup}{\coq version} is almost a direct transcription:
\begin{vuplisting}
Definition fwd_or_clause_base G Δ s c : edelta := let GΔ := edb :+: Δ in
 let T = [set gr σ c.headc | σ in \bigcup_(b <- c.bodyc) match_body G b ] in
 Δ{s $\to$ GΔ s $\ominus$ T}
\end{vuplisting}
in the first line, we use a set comprehension to build the set of
ground facts corresponding to the consequence operator. Note that
\C|\bigcup_(x <- X)| is the \coq notation for $\bigcup\limits_{x \in X}$ and
that \C|gr $\sigma$ head| denotes the application of the substitution $\sigma$ to 
the input head. The second line updates the resulting $\up$, using the operator for
\emph{modification} (see Def.~\ref{def:gmod}) and that for \emph{graph difference} ($\ominus$).
We remark that this base operator will re-derive all the facts, as it
does not make use of incrementality information.
\subsection{Incremental Delta-Join Maintenance}\label{sec:engine:delta-join}
%
%
%
Given a graph \g, a program $\Pi$, and updates $\up$, the engine in
Sec.~\ref{sec:engine:base} \emph{non-incrementally} maintains the
top-level view of $\Pi$. However, the engine is unable to \emph{reuse}
and \emph{adjust} previously computed maintenance information.  This
is makes it especially inefficient when few nodes are added to an
otherwise high-cardinality graph.

To remedy this situation, we would like to extend our algorithm so that it
can take into account the information of previously computed
models. The key idea is to restrict \emph{matching} to graph updates.
For example, let $V$ be a materialized view, defined as a simple
\emph{join}, in our case, as the path over two base edges, $r$ and
$s$, i.e. $V(X,Y) \leftarrow r(X,Z), s(Z,Y)$. We abbreviate this as
$V = r \bowtie s$. Given \emph{base deltas}, $\LD{r}$
and $\LD{s}$, we can compute the \emph{view delta} as
$\up V = (\LD{r} \bowtie s) \cup (r \bowtie \LD{s})
\cup (\LD{r} \bowtie \LD{s})$, or, after factoring, as
$\LD{V} = (\LD{r} \bowtie s) \cup (\LF{r} \bowtie \LD{s})$, where $\LF{r} = r \cup \LD{r}$.  Hence,
$\LD{V} = \LD{V}_{1} \cup \LD{V}_{2}$, with $\LD{V}_{1}$ and
$\LD{V}_{2}$ computable via the \emph{delta clauses}:
$\delta_{1}: \LD{V}_{1} \leftarrow \LD{r}(X,Z), \LD{s}(Z,Y)$ and
$\delta_{2}: \LD{V}_{2} \leftarrow \LF{r}(X,Z), \LD{s}(Z,Y)$.

Generalizing, for a database $\mathcal{G}$ and a purely additive update
$\up$, we can determine the \emph{view delta}
$\LD{V}[\g; \up]$, i.e., the set of facts
such that
$V[\appd{\g}{\up}] = V[\g] \cup \LD{V}[\g; \up]$.
%
\begin{definition}[Delta Program] Let $V$ be a view defined by
  $V \leftarrow L_{1}, \ldots, L_{n}$.  The \emph{delta program}
  $\delta(V)$ is $\{\delta_{i} \mid i \in [1,n]\}$. Each \emph{delta
    clause} $\delta_{i}$ has the form
  $V \leftarrow L_1, \ldots, L_{i-1}, \LD{L}_i, \LF{L}_{i+1}, \ldots, \LF{L}_{n}$, where:
  $L^{\nu}_{j}$ marks that we match $L_j$ against atoms in $\mathcal{G} \cup \Delta{G}$ with the same symbol as $L_j$
  and $\LD{L}_{j}$ marks that we match $L_j$ against atoms in $\Delta\g$ with the same symbol as $L_j$.
\end{definition}
We revisit Example~\ref{ex:fd} to illustrate the computation of incremental view updates.\\
\begin{example}[Detectable Frauds]\label{ex:df}
  Consider the transaction configuration in
  Fig.~\ref{fig:sub1}, marking suspect transactions. A
  suspect transaction between $X$ and $Y$ is \emph{detectable}, if
  there exists an entity that monitors both $X$ and $Y$. We can
  compute all detectable suspect transactions with the RQ:
  \[
    \begin{array}{lll}
      \mathbf{d}etectable(X,Y) & \leftarrow & \mathbf{s}uspect(X,Y), \mathbf{m}onitors(Z,X), \mathbf{m}onitors(Z,Y)
    \end{array}
  \]
  These are
  $detectable = \{(V_{6}, V_{0}), (V_{3}, V_{0}))\}$. When \emph{updating} the previous graph to the one in Figure~\ref{fig:dfp2}, we have:
  $detectable^{\nu} = \{(V_{6}, V_{0}), (V_{3}, V_{0}), \mathbf{(V_{0}, V_{2})}, \mathbf{(V_{2}, V_{0})}, \mathbf{(V_{0}, V_{5})}\}$.
  The delta update
  $\LD{detectable} = \{\mathbf{(V_{0}, V_{2})}, \mathbf{(V_{2},
    V_{0})}, \mathbf{(V_{0}, V_{5})}\}$ can be \emph{incrementally} computed with 
  the program $\Pi_{\Delta} = \delta_{1} \cup \delta_{2} \cup \delta_{3}$, as follows: $\delta_{1} =  \emptyset$, $\delta_{2} = \{\mathbf{(V_{2}, V_{0})}\}$, and
  $\delta_{3} = \{\mathbf{(V_{0}, V_{2})}, \mathbf{(V_{0}, V_{5})}\}$.
  Indeed, $\LD{detectable} = \LF{detectable} \setminus detectable$.
  \vspace{-2mm}
  \begin{figure}[h!]
    \centering
    \subfloat[Initial Graph]{
      \begin{tikzpicture}[scale=1, every node/.style={scale=0.7}]\label{fig:sub1}
  \tikzstyle{every state}=[fill=none,draw=black,text=black, minimum size=0.5mm, inner sep=0.8pt]
  \node[state]         (V0) at (0,0)      {$V_{0}$};
  \node[state]         (V1) at (-1,1)     {$V_{1}$};
  \node[state]         (V2) at (1,1)      {$V_{2}$};
  \node[state]         (V3) at (1,0)      {$V_{3}$};
  \node[state]         (V4) at (1,-1)     {$V_{4}$};
  \node[state]         (V5) at (-1,-1)    {$V_{5}$};
  \node[state]         (V6) at (-1,0)     {$V_{6}$};
  \path (V1) edge[->,above, sloped]      node {\small{\textbf{m}}}  (V0)
        (V1) edge[->,above, sloped]      node {\small{\textbf{m}}}  (V6)
        (V0) edge[->,above, sloped]      node {\small{\textbf{s}}}  (V2)
        (V2) edge[->,above, sloped]      node {\small{\textbf{s}}}  (V3)       
        (V6) edge[->,above, sloped]      node {\small{\textbf{s}}}  (V0)
        (V3) edge[->,above, sloped]      node {\small{\textbf{s}}}  (V0)
        (V5) edge[->,above, sloped]      node {\small{\textbf{s}}}  (V6)
        (V0) edge[->,above, sloped]      node {\small{\textbf{s}}}  (V5)
        (V4) edge[->,above, sloped]      node {\small{\textbf{m}}}  (V0)
        (V4) edge[->,above, sloped]      node {\small{\textbf{m}}}  (V3);
  \end{tikzpicture}
  }
  \qquad
  \subfloat[Updated Graph]{
  \begin{tikzpicture}[scale=1, every node/.style={scale=0.7}]\label{fig:dfp2}
  \tikzstyle{every state}=[fill=none,draw,text=black, minimum size=0.5mm, inner sep=0.8pt]
  \node[state]         (V0) at (0,0)      {$V_{0}$};
  \node[state]         (V1) at (-1,1)     {$V_{1}$};
  \node[state]         (V2) at (1,1)      {$V_{2}$};
  \node[state]         (V3) at (1,0)      {$V_{3}$};
  \node[state]         (V4) at (1,-1)     {$V_{4}$};
  \node[state]         (V5) at (-1,-1)    {$V_{5}$};
  \node[state]         (V6) at (-1,0)     {$V_{6}$};
  \path (V1) edge[->,above, sloped]      node {\small{\textbf{m}}}  (V0)
        (V1) edge[->,above, sloped]      node {\small{\textbf{m}}}  (V6)
        (V0) edge[->,above, bend left=10, sloped] node {\small{\textbf{s}}}  (V2)
        (V2) edge[->,above, sloped]      node {\small{\textbf{s}}}  (V3)       
        (V6) edge[->,above, sloped]      node {\small{\textbf{s}}}  (V0)
        (V3) edge[->,above, sloped]      node {\small{\textbf{s}}}  (V0)
        (V5) edge[->,above, sloped]      node {\small{\textbf{s}}}  (V6)
        (V0) edge[->,above, sloped]      node {\small{\textbf{s}}}  (V5)
        (V4) edge[->,above, sloped]      node {\small{\textbf{m}}}  (V0)
        (V4) edge[->,above, sloped]      node {\small{\textbf{m}}}  (V3)
        (V1) edge[->,above, sloped, thick]  node {\small{\textbf{m}}}  (V2)
        (V4) edge[->,above, sloped, thick]  node {\small{\textbf{m}}}  (V5)
        (V2) edge[->,below, bend left=10, sloped, thick]  node {\small{\textbf{s}}}  (V0)
        ;
  \end{tikzpicture}
  }
  \\
  \subfloat[Delta Program for Detectable Frauds]{\small
    $
    \begin{array}{lll}
      \delta_{1}: \quad \LD{detectable}(X,Y) & \leftarrow & \LD{\mathbf{s}uspect}(X,Y), \mathbf{m}onitors(Z,X), \mathbf{m}onitors(Z,Y) \\
      \delta_{2}: \quad \LD{detectable}(X,Y) & \leftarrow & \LF{\mathbf{s}uspect}(X,Y), \LD{\mathbf{m}onitors}(Z,X), \mathbf{m}onitors(Z,Y) \\
      \delta_{3}: \quad \LD{detectable}(X,Y) & \leftarrow & \LF{\mathbf{s}uspect}(X,Y), \LF{\mathbf{m}onitors}(Z,X), \LD{\mathbf{m}onitors}(Z,Y) \\
    \end{array}
    $
  }
  \label{fig:dfp}
  \caption{Detectable Frauds}
\end{figure}
\end{example}

\subsection{The $\Delta$-Engine for RD Evaluation and Incremental View Maintenance}
\label{sec:engine:delta}
We now present the incremental version of the clause evaluation
operator defined in Sec.~\ref{sec:engine:base}. We follow
Sec.~\ref{sec:engine:delta-join} and modify base matching to take into
account $\Delta$-clauses and programs. Thus, for each body to be
processed incrementally, we generate a \emph{body mask}, placing a tag
--- $\{\mathbf{B},\mathbf{D},\mathbf{F}\}$ --- on each body literal, which 
indicates whether matching should proceed against the base
interpretation, against the update, or against both.
The \emph{incremental atom matching} operator $\matchda$
(\inlinelink[\#match_delta_atoms]{VUP.vup}{\C|match_delta_atoms|}) is defined as:
\begin{equation*}
  \matchda(a) = (\mathsf{if}~m \in \bra{\mathbf{B},\mathbf{F}}~\mathsf{then}~\matcha[\g](a) ~\mathsf{else}~\emptyset) \cup
                (\mathsf{if}~m \in \bra{\mathbf{D},\mathbf{F}}~\mathsf{then}~\matcha[\up](a)~\mathsf{else}~\emptyset)
\end{equation*}
thus, base matching is called with the instance corresponding
to the atom's tag.

Body $\Delta$-matching, $\matchdb$ (\inlinelink[\#match_delta_body]{VUP.vup}{\C|match_delta_body|}) takes as an
input a \emph{body mask}, that is to say, a list of tag-annotated
literals.  A function \inlinelink[\#body_mask]{VUP.vup}{\C|body_mask|} generates the set $B_\Delta$ of
``decorated'' literals. Generic syntax is extremely helpful here to
avoid duplication and to help state mask invariants in an elegant
way. \C|body_mask| follows the \emph{diagonal factoring} described
below, where each row corresponds to an element of $B_\Delta$:
\begin{equation*}
  \begin{bmatrix}
     \MD{L_1} & \MF{L_2} & \ldots & \MF{L_{n-1}} & \MF{L_n} \\
     \MB{L_1} & \MD{L_2} & \ldots & \MF{L_{n-1}} & \MF{L_n} \\
     \hdotsfor{5} \\
     \MB{L_1} & \MB{L_2} & \ldots & \MB{L_{n-1}} & \MD{L_n} \\
   \end{bmatrix}
\end{equation*}
%
The last piece to complete the incremental engine is the top-level
clausal maintenance operator. This part of the engine is significantly
more complex than its \emph{base} counterpart, as it must take into
account which incrementality heuristics to apply:
\begin{definition}[Incremental Clausal Maintenance Operator]
  The $\iTcl{s}(\up)$ operator for incremental clausal maintenance
  (\inlinelink[\#fwd_or_clause_delta]{VUP.vup}{\C|fwd_or_clause_delta|}) acts in
  two cases. If $s \notin \supp$, or $\up$ contains deletions for any
  of the literals in the body of $\Pi(s)$, it uses the base operator
  $\iTBcl{s}(\appd{\g}{\up})$ --- as we either cannot reuse the
  previous model or cannot support deletions through our incremental
  strategy.  Otherwise, the operator will generate
  $B_\up = \text{\C|body_mask|}(B)$, for each of the bodies $B$, and
  return $\bigcup_{B_m \in B_\up} \matchdb(B_m)$.
\end{definition}

%


\section{\rdlong Evaluation: Certified Soundness}\label{sec:sound}
\label{sec:proof}
We now summarize the main technical points of the mechanized proof
developed in \coq.
A key to effective mechanized proof development is the definition of
the proper high-level concepts and theories; unfortunately we lack the
space here to describe all the definitions used in our mechanized
development in detail, so we highlight the main result, that proves
the soundness of the engine, and we briefly survey the two core
theories for \emph{stratification} and \emph{incrementality}.
A few auxiliary results are described in~\ref{sec:proof_details}.

\subsection{Stratification Conditions}
\begin{definition}[Stratified Programs]\label{sec:slp}
  A key precondition for the \emph{soundness} of our engine is
  program stratification.
  A program $\Pi$ is \emph{stratified}, if there exists a mapping
  $\sigma : \gsig \rightarrow [1,n]$ such that, for all $s$ in
  $\gsig$, the $\Pi(s)$ clause $(t_1,t_2) \leftarrow B$ satisfies:
  ${\max\limits_{r \in \sym(B)}} \sigma(r) < \sigma(s)$, where
  $\sym$ returns the set of symbols occurring in $B$.
  We then call $\sigma$ a \emph{stratification} of $\Pi$. In \coq,
  we encode stratification using a list of symbols
  and a predicate \inlinelink[\#is_strata_rec]{VUP.vup}{\C|is_strata|} that
  recursively checks each uninspected symbol against an accumulator.
  This choice of representation is practical as it will guide model
  computation in the engine.
\end{definition}
\begin{definition}[Well-formed Program Slices]\label{sec:wf_slice}
  In order to reason about stratified satisfaction, we need a
  strengthened notion of well-formedness stating that a program is
  closed w.r.t. a symbol set. A symbol set $\sset$ is a
  \emph{well-formed slice} of $\Pi$ if, for all $s$ in $\sset$,
  $\sym(\Pi(s)) \subseteq \sset$.
\end{definition}
We establish that the engine operates over well-formed slices, which
allows us to isolate reasoning about the current iteration, see the
appendix for more technical details.
%



\subsection{IGVM-Engine Characterization Result}
\label{sec:djcoq}
Let $\Pi$ be a \emph{safe} \rd program, $\strl$ and $\strg$,
symbol sets corresponding  the ``extensional'' (already processed)
and ``intensional'' (to be processed) strata, $\g$ a graph
instance, and $\up$ an update. Then we establish:

\begin{theorem}[IGVM-Engine Soundness]
  Assume (H1) $\g$ is a model of the program for $\supp$,
  $\csat{\supp}{\Pi}$, (H2) $\strl$ is a \emph{well-formed slice} of
  $\Pi$, (H3) $\up$ only contains information up to the processed
  strata, $\sym(\up) \subseteq \strl$, (H4) $\strl, \strg$ are a
  \emph{stratification} of $\Pi$, and (H5) the currently model is
  sound, $\dsat{\strl}{\Pi}$ then, the engine --- implementing the
  maintenance operator $\iTp(\strl,\strg,\up)$ 
  --- outputs an update $\up_O$, such that
  $\dsat[\up_O]{\gsig}{\Pi}$ holds.
\begin{proof}
  The proof is a consequence of the soundness of the \emph{incremental
    clausal maintenance operator}
  $\iTcl{s}$~(Lem.~\ref{lemma:fwdorclP}). We proceed by
  \emph{induction} on $\strg$. The \emph{base case} follows from
  $\strl = \gsig$, as $\dsat{\strl}{\Pi}$ holds by assumption.
  For the \emph{inductive case}, let $\strg \equiv \{s\} \cup \strg'$ and
  $C$ be the clause $\Pi(s)$. Given a set of symbols $S$ and an update
  $\up_O$, such that $\dsat[\up_O]{S}{\Pi}$, the \emph{induction hypothesis}
  (\textbf{IH}) ensures that $\dsat[\up_O]{S \cup \strg'}{\Pi}$.

  Now, we need to prove that
  $\dsat[\iTp(\uset{s}{\strl},\strg',\iTcl{s}(\up))]{\strl \cup \uset{s}{\strg'}}{\Pi}$.
  The conclusion results from instantiating (\textbf{IH}) with
  $S = \uset{s}{\strl}$ and $\up_O = \iTcl{s}(\up)$. To this end, we need to
  first establish that $\dsat[\iTcl{s}(\up)]{\uset{s}{\strl}}{\Pi}$. This ensues
  from clause maintenance soundness and from modularity of model satisfaction.

  The rest of preconditions needed by \textbf{IH} hold, as follows. From H4,
  since $(\uset{s}{\strg'}) \cup \strl$ \emph{stratifies} $\Pi$, we have that
  $\strg' \cup \uset{s}{\strl}$ also stratifies $\Pi$. Moreover,
  $s \notin \strl$, and $\sym(C) \subseteq \strl$. From H2, we have that
  $\strl$ is a \emph{well-formed slice} of $\Pi$, which, together with
  $\sym(C) \subseteq \strl$, proves $\{s\} \cup \strl$ is
  a well-formed slice. From H3, we know that $\sym(\Delta) \subseteq 
  \strl$. The auxiliary \inlinelink[\#fwd_or_clause]{VUP.vup}{\C|supp$\Delta$_fwd_or_clause|} lemma ensures that
  $\sym(\up_O) \subseteq \uset{s}{\sym(\up)}$; it then follows,
  by transitivity, that $\sym(\up_O) \subseteq \{s\} \cup \strl$.
  The \coq proof is about 25 lines long and comparable to this
  text-based version in size. The first line sets up the induction,
  with the rest of the proof consisting in the instantiation of the
  proper lemmas. The statement of the theorem itself takes 7 lines for
  the preconditions (1 per line) plus an additional line for the
  conclusion: \C|ssTp (edb :+: $\Delta_O$) ($\sset$ $\cup$ $\strl$) $\Pi$|.
%
\end{proof}
\end{theorem}
%




\section{Experimental Analysis}\label{sec:exp}
\label{sec:eval}
In this section, we present the experimental validation of our certified 
engine on realistic graph databases. Our empirical analysis aims 
to confirm that \emph{incremental view maintenance (IVM)} is 
more beneficial than \emph{full view materialization (FVM)}, corresponding to recomputing the view from scratch, 
at each modification of the underlying data. The comparison has been 
established by computing the runtimes on the same engine.  

As it is common practice in the
verification community, our extracted engine has been obtained 
through the mechanism of \emph{program extraction}~\cite{Letouzey}, starting 
from our underlying \coq formalization. 
%
Assuming that \coq extraction is semantics preserving, also the underlying premise of our engine, 
the \emph{correctness} of the obtained OCaml 
engine is readily guaranteed by the \coq specification we provided. This is a reasonable
assumption, made by past verified tools, such as~\cite{DBLP:journals/cacm/Leroy09}.

Our tests have been performed on a Intel Core i7 vPro G6 laptop, with 16GB
RAM, running Ubuntu 17.10 64 bit, and OCaml 4.06.0.

For our experimental analysis, we generated synthetic datasets and query workloads using gMark~\cite{gmark}, 
which allowed us to encode, two state-of-the-art benchmarks: WD, the Waterloo SPARQL Diversity Test Suite (Wat-Div)~\cite{WD}, and SNB, the LDBC Social Network Benchmark~\cite{LDBC}. 
These graphs -- henceforth denoted by $\g$ --
are diverse in terms of their density (increasing from SNB to WD) and of their in-degree and 
out-degree distributions~\cite{gmark}. They represent two extreme cases to 
be considered in benchmarking graph database engines. Each schema size $|supp(\g)|$ is fixed at 82 and 27 predicates, for WD, respectively, SNB.

%
Based on this, we generate graph instances and companion query workloads, such that $|\g| = 1K$ nodes and
$|\mathcal{W}| = 10$ queries (queries represent views in our setting). 
Note that the gMark-generated queries are UC2RPQs -- a notable subset of Regular Queries (RQs)~\footnote{To 
the best of our knowledge, there currently is no practical benchmark capable of generating query workloads over the full fragment of Regular Datalog studied in this paper. 
The generation of graph query workloads for UC2RPQs has indeed been proved to be NP-complete~\cite{gmark}.}. This should not be considered a restriction,
as these queries already let us stress (for view maintenance) the navigational part of our engine and use the recursion in the form of Kleene-star -- a 
bottleneck in many practical graph query engines (such as Neo4J)~\cite{gmark}. 

In order to build the deltas necessary for incremental view maintenance, we sampled the original graph instance $\g$ by \emph{support size}, i.e., by considering arbitrary subgraphs, whose number of symbols
represent certain fractions of those in $\g$. We call this method \emph{symbol-based sampling}. Concretely, we retained a varying percentage 
$\rho_{supp} \in \{0.05, 0.1, 0.15, 0.2, 0.25\}$ from $supp(\g)$. Next, we took all corresponding 
instance facts with symbols in $\rho_{supp}$ to be our \emph{bulk insertions}, $\Delta_{+}$, 
where $\frac{|supp(\Delta^{+})|}{|supp(\g)|} = \rho_{supp}$.
These bulk insertions correspond to adding the subgraphs of $\g$, whose sets of symbols,
sampled from $supp(\g)$, reflect the $\rho_{supp}$ ratio. We consider 
$\g' = \g \setminus \Delta_{+}$ 
to be our \emph{base instances}, i.e., the initial sets of facts to be processed by our engine.

Since we relied on symbol-based sampling, 
the actual sizes of deltas ($|\Delta_{+}|$) vary depending on the content of 
$\g$. We denote the percentage capturing the relative sizes of the bulk inserts with respect to those of the base instances,
as $\rho = \frac{|\Delta_{+}|}{|\mathcal{G'}|} * 100$ and report its actual values in the second column of each of the Tables~\ref{fig:expwd} and~\ref{fig:expsnb}.


Next, we evaluated each query in a workload $\mathcal{W}$ over $\g'$ and materialized the 
resulting views. Upon updating $\g'$ with $\Delta_{+}$, we compared the 
\emph{average timings} for \emph{incremental view maintenance} (IVM) and \emph{full view recomputation} (FVM), over all view materializations, in each of the workloads.
We summarize the results in Table~\ref{fig:expwd} and Table~\ref{fig:expsnb}.

%
We observe that the \emph{absolute time gain (ms)} of our engine running IVM with respect to it running FVM, i.e., $\textbf{Time Gain} = \text{FVM} - \text{IVM}$, is \emph{always positive} 
and that the \emph{relative ratio gain (\%)}, i.e., $\textbf{Ratio Gain} = 100 - \frac{100*\text{IVM}}{\text{FVM}}$, is always better for sparser graphs. As expected, our engine works best on bulk updates involving very small individual symbol updates, as these types of updates
are targeted by delta join matching. Indeed, the complexity of our delta join depends on how many matchings have to be computed. Note that the lower the 
sparsity of the underlying graph, the less matches we have and the faster our engine is. This explains why the runtimes over SNB (less dense) are comparatively much 
better than the ones over WD (very dense). 
\begin{table}[h!]
\begin{tabular}{|c|c|c|c|c|c|}  \cline{1-6} 
 $\mathbf{\rho_{supp}}$    & $\mathbf{\rho}$ &\textbf{FVM}    & \textbf{IVM}   & \textbf{Time Gain}  & \textbf{Ratio Gain} \\ \cline{1-6} 
 0.05       & 1.4\% & 558.7 & 484.75 & 73.95 &  13.23\%   \\ \cline{1-6} 
 0.1        & 3.67\% & 561.89 & 472.7 & 89.19 &  15.87\%   \\ \cline{1-6} 
 0.15       & 17.93\% & 562.67 & 475.96 & 86.71 &  15.41\% \\ \cline{1-6} 
 0.2        & 9.7\% & 562.13 & 476.4 & 85.73 &  15.25\%   \\ \cline{1-6} 
 0.25       & 18.26\% & 563.4 & 482.64 & 80.76 &  14.33\% \\ \cline{1-6} 
\end{tabular} 
\vspace{2mm}
\caption{Avg. $\mathcal{W}_{\twd}$ Runtimes (ms) for Varying Support Update Size ($\rho_{supp}$) }
\label{fig:expwd}
\end{table}
\vspace{-6mm}
\begin{table}[h!]
\begin{tabular}{|c|c|c|c|c|c|} \cline{1-6}
 $\mathbf{\rho_{supp}}$  & $\mathbf{\rho}$ & \textbf{FVM} & \textbf{IVM}  & \textbf{Time Gain}  & \textbf{Ratio Gain} \\ \cline{1-6}
 0.05       & 10.89\% &18.75 & 10.88 & 7.87 & 41.97\%  \\ \cline{1-6}
 0.1        &  19.3\% &17.77 & 10.55 & 7.22 &  40.63\%  \\ \cline{1-6}
 0.15       &  10.77\% &17.55 & 11.68 & 5.82 &  33.25\% \\ \cline{1-6}
 0.2        &  26.09\% &17.17 & 11.71 & 5.46 &  31.79\%  \\ \cline{1-6}
 0.25       &  28.34\% &14.71 & 11 & 3.71 &  25.22\% \\ \cline{1-6}
\end{tabular} 
\vspace{2mm}
\caption{Avg. $\mathcal{W}_{\tsnb}$ Runtimes (ms) for Varying Support Update Size ($\rho_{supp}$) }
\label{fig:expsnb}
\end{table}

\section{Related Work}\label{sec:related}
\label{sec:rel}
To the best of our knowledge, no verified graph query or IVM engine exists, 
which we both design and mechanize in this paper, using Regular Datalog. 
Bounded~\footnote{The theory of bounded computational complexity for dynamic
graph problems~\cite{RamalingamR96} considers the cost of incremental computation as a
polynomial function of the input and output changes.} incremental graph computation has been addressed
in~\cite{FanHT17} and shown beyond reach already with Regular Path Queries (RPQs), a restricted navigational RQ subset.
The paper's idea is to incrementalize the bulk RPQ evaluation,
by leveraging NFAs and auxiliary structures on large-scale graphs. We
focus instead on the \emph{verification} of forward-chaining-based IVM for
the more expressive RQ graph query fragment. 
Although recent work has addressed certifying SQL
semantics~\cite{ChuWCS17}, by proving the semantic preservation of
rewriting rules in SQL query optimizers for relational data, such
a mechanization is not applicable to the graph-data setting, where the key
data model component is no longer a tuple, but a \emph{path} (connecting
pairs of graph nodes). To this end, our Coq development is based on
the standard connectivity notion from the Mathematical Components library.

Similarly, verified frameworks for the relational
data model and nested relational algebra query
compilers~\cite{BenzakenCD14,AuerbachHMSS17} are fairly orthogonal to our work. The
optimization issues around RQ evaluation have never been \emph{formally
addressed} in the database literature and, even for the simple 
UC2RPQ class, current graph database engines perform poorly (see~\cite{gmark}). 
Even though our goal is not to provide a RQ optimizer, we touch base with some simple optimizations. 
These are clause normalization, a lightweight indexing mechanism (leveraging graph edges in the definition of \emph{supports}),
and our \emph{incremental supported satisfaction} definition. This leads to a more elegant framework for reasoning about incremental properties. 

Despite RD not being fully recursive, our engine handles \emph{stratified evaluation} in a Datalog style. 
Focusing on linear recursion~\cite{Jagadish87} is indeed sufficient for our purposes, as it allows us to build a \emph{RQ-specific engine}
that inherently handles transitive closures. By limiting recursion, we can express \emph{graph recursive} queries (RQs) that are efficiency prone, 
being highly parallelizable~\cite{GreenLawHR95}. 
The work in~\cite{BCD17} presents the SSReflect certification of a stratified static Datalog toy engine, implementing the bottom-up heuristic. 
While it supports Datalog's full recursion, \emph{this is actually a bottleneck} even for non-verified graph query engines~\cite{gmark}; also, it 
does not handle graph updates and IVM. 

Finally, efficient Datalog engines have been designed in the last two
decades, such as
DLV2~\cite{AlvianoCDFLPRVZ17} and LogicBlox~\cite{ArefCGKOPVW15}. 
Our proposed methodology can be implemented on top of such advanced engines, to combine their efficiency with our certified mechanization.
We hope that our work paves the way for a future interplay of the various optimized heuristics and implemented semantics
for Datalog, with its comprehensively verified evaluation.
While the \coq extraction mechanism is mature and well-tested, we plan to 
integrate the recent advances
on the trusted extraction and
compilation~\cite{Anand2016CertiCoqAV,Mullen:2018:MCE:3176245.3167089}
of \coq code into our framework.
%
%

\section{Conclusion and Perspectives}
\label{sec:concl}
We propose a \coq formal library for \emph{certified incremental graph query evaluation and view maintenance} in the \rdlong fragment.
It consists of 1062 lines of definitions, specifying our mechanized theory, and 734 lines of proofs, establishing the central \emph{soundness guarantee}.
Our mechanized specification builds on a library fine-tuned for the computer-aided theorem proving of finite-set theory results. We take advantage from this, by
giving a high-level, \emph{mathematical} representation of core engine components. This leads to composable lemmas that boil down
to set theoretic statements and, ultimately, to a condensed development, avoiding the proof-complexity explosion characteristic
of formal verification efforts. Moreover, we managed to \emph{extract}
a runnable engine exhibiting performance gains versus the
non-incremental approach on realistic graph database benchmarks.
Our foundational approach shows it is promising to combine logic programming and proof assistants, such as \coq, to give
certified specifications for both a uniform graph query language and its evaluation. We plan to mechanize more optimized heuristics, particularly
for efficiently handling joins, and to integrate custom algorithms, such as Tarjan, for transitive closure computation.

\paragraph{Acknowledgments:} We would like to thank the anonymous referees
and Pierre Jouvelot for their very useful comments and
feedback. S.Dumbrava was funded by the Datacert project, ANR-15-CE39-0009, and by ANR-11-IDEX-0007.

\bibliographystyle{acmtrans}
{\small\bibliography{paper-bib}}

\begin{thebibliography}{}

\bibitem[\protect\citeauthoryear{Abiteboul, Hull, and Vianu}{Abiteboul
  et~al\mbox{.}}{1995}]{Alice}
{\sc Abiteboul, S.}, {\sc Hull, R.}, {\sc and} {\sc Vianu, V.}, Eds. 1995.
\newblock {\em Foundations of Databases: The Logical Level\/}, 1st ed.
\newblock Addison-Wesley Longman Publishing Co., Inc., Boston, MA, USA.

\bibitem[\protect\citeauthoryear{Aluç, Hartig, Özsu, and Daudjee}{Aluç
  et~al\mbox{.}}{2014}]{WD}
{\sc Aluç, G.}, {\sc Hartig, O.}, {\sc Özsu, M.~T.}, {\sc and} {\sc Daudjee,
  K.} 2014.
\newblock Diversified stress testing of {RDF} data management systems.
\newblock In {\em The Semantic Web (ISWC 2014)}, {P.~Mika}, {T.~Tudorache},
  {A.~Bernstein}, {C.~Welty}, {C.~Knoblock}, {D.~Vrande{\v{c}}i{\'{c}}},
  {P.~Groth}, {N.~Noy}, {K.~Janowicz}, {and} {C.~Goble}, Eds. {LNCS}, vol.
  8796. Springer International Publishing, Cham, 197--212.

\bibitem[\protect\citeauthoryear{Alviano, Calimeri, Dodaro, Fusc{\`{a}}, Leone,
  Perri, Ricca, Veltri, and Zangari}{Alviano
  et~al\mbox{.}}{2017}]{AlvianoCDFLPRVZ17}
{\sc Alviano, M.}, {\sc Calimeri, F.}, {\sc Dodaro, C.}, {\sc Fusc{\`{a}}, D.},
  {\sc Leone, N.}, {\sc Perri, S.}, {\sc Ricca, F.}, {\sc Veltri, P.}, {\sc
  and} {\sc Zangari, J.} 2017.
\newblock The {ASP} system {DLV2}.
\newblock In {\em Logic Programming and Nonmonotonic Reasoning {LPNMR} 2017}.
  215--221.

\bibitem[\protect\citeauthoryear{Anand, Appel, Morrisett, Paraskevopoulou,
  Pollack, B{\'e}langer, Sozeau, and Weaver}{Anand
  et~al\mbox{.}}{2017}]{Anand2016CertiCoqAV}
{\sc Anand, A.}, {\sc Appel, A.~W.}, {\sc Morrisett, G.}, {\sc Paraskevopoulou,
  Z.}, {\sc Pollack, R.}, {\sc B{\'e}langer, O.~S.}, {\sc Sozeau, M.}, {\sc
  and} {\sc Weaver, M.} 2017.
\newblock Certicoq: A verified compiler for {Coq}.
\newblock In {\em {CoqPL} 2017: The 3rd International Workshop on {Coq} for
  Programming Languages}.

\bibitem[\protect\citeauthoryear{Angles, Arenas, Barcel{\'{o}}, Hogan, Reutter,
  and Vrgoc}{Angles et~al\mbox{.}}{2017}]{AnglesABHRV17}
{\sc Angles, R.}, {\sc Arenas, M.}, {\sc Barcel{\'{o}}, P.}, {\sc Hogan, A.},
  {\sc Reutter, J.~L.}, {\sc and} {\sc Vrgoc, D.} 2017.
\newblock Foundations of modern query languages for graph databases.
\newblock In {\em {ACM} Comput. Surv.} Vol.~50. 68:1--68:40.

\bibitem[\protect\citeauthoryear{Aref, ten Cate, Green, Kimelfeld, Olteanu,
  Pasalic, Veldhuizen, and Washburn}{Aref et~al\mbox{.}}{2015}]{ArefCGKOPVW15}
{\sc Aref, M.}, {\sc ten Cate, B.}, {\sc Green, T.~J.}, {\sc Kimelfeld, B.},
  {\sc Olteanu, D.}, {\sc Pasalic, E.}, {\sc Veldhuizen, T.~L.}, {\sc and} {\sc
  Washburn, G.} 2015.
\newblock Design and implementation of the {LogicBlox} system.
\newblock In {\em Proceedings of {ACM} {SIGMOD}}. 1371--1382.

\bibitem[\protect\citeauthoryear{Auerbach, Hirzel, Mandel, Shinnar, and
  Sim{\'e}on}{Auerbach et~al\mbox{.}}{2017}]{AuerbachHMSS17}
{\sc Auerbach, J.~S.}, {\sc Hirzel, M.}, {\sc Mandel, L.}, {\sc Shinnar, A.},
  {\sc and} {\sc Sim{\'e}on, J.} 2017.
\newblock Handling environments in a nested relational algebra with combinators
  and an implementation in a verified query compiler.
\newblock In {\em Proceedings of the 2017 ACM International Conference on
  Management of Data}. SIGMOD '17. ACM, New York, NY, USA, 1555--1569.

\bibitem[\protect\citeauthoryear{Bagan, Bonifati, Ciucanu, Fletcher, Lemay, and
  Advokaat}{Bagan et~al\mbox{.}}{2017}]{gmark}
{\sc Bagan, G.}, {\sc Bonifati, A.}, {\sc Ciucanu, R.}, {\sc Fletcher, G.
  H.~L.}, {\sc Lemay, A.}, {\sc and} {\sc Advokaat, N.} 2017.
\newblock {gMark}: Schema-driven generation of graphs and queries.
\newblock {\em {IEEE} Transactions on Knowledge and Data Engineering\/}~{\em
  29,\/}~4 (April), 856--869.

\bibitem[\protect\citeauthoryear{Benzaken, Contejean, and Dumbrava}{Benzaken
  et~al\mbox{.}}{2014}]{BenzakenCD14}
{\sc Benzaken, V.}, {\sc Contejean, E.}, {\sc and} {\sc Dumbrava, S.} 2014.
\newblock A {Coq} formalization of the relational data model.
\newblock In {\em Proceedings of the 23rd European Symposium on Programming
  Languages and Systems - Volume 8410}. Springer-Verlag New York, Inc., New
  York, NY, USA, 189--208.

\bibitem[\protect\citeauthoryear{Benzaken, Contejean, and Dumbrava}{Benzaken
  et~al\mbox{.}}{2017}]{BCD17}
{\sc Benzaken, V.}, {\sc Contejean, E.}, {\sc and} {\sc Dumbrava, S.} 2017.
\newblock Certifying standard and stratified {Datalog} inference engines in
  {SSReflect}.
\newblock In {\em Interactive Theorem Proving}. {LNCS}, vol. 10499. Springer
  International Publishing, 171--188.

\bibitem[\protect\citeauthoryear{Beyhl and Giese}{Beyhl and
  Giese}{2016}]{BeyhlG16}
{\sc Beyhl, T.} {\sc and} {\sc Giese, H.} 2016.
\newblock Incremental view maintenance for deductive graph databases using
  generalized discrimination networks.
\newblock In {\em {GaM@ETAPS}}. {EPTCS}, vol. 231. 57--71.

\bibitem[\protect\citeauthoryear{Cai, Giarrusso, Rendel, and Ostermann}{Cai
  et~al\mbox{.}}{2014}]{Cai:2014:TCH:2594291.2594304}
{\sc Cai, Y.}, {\sc Giarrusso, P.~G.}, {\sc Rendel, T.}, {\sc and} {\sc
  Ostermann, K.} 2014.
\newblock A theory of changes for higher-order languages: Incrementalizing
  {$\lambda$}-calculi by static differentiation.
\newblock In {\em Proceedings of the 35th ACM SIGPLAN Conference on Programming
  Language Design and Implementation}. PLDI '14. ACM, New York, NY, USA,
  145--155.

\bibitem[\protect\citeauthoryear{??}{{Cayley}}{}]{Cayley}
{Cayley}.
\newblock \url{https://github.com/cayleygraph/cayley} (visited: 2018-02).

\bibitem[\protect\citeauthoryear{Ceri, Gottlob, and Tanca}{Ceri
  et~al\mbox{.}}{1989}]{Ceri}
{\sc Ceri, S.}, {\sc Gottlob, G.}, {\sc and} {\sc Tanca, L.} 1989.
\newblock What you always wanted to know about {Datalog} (and never dared to
  ask).
\newblock {\em {IEEE} Transactions on Knowledge and Data Engineering\/}~{\em
  1,\/}~1, 146--166.

\bibitem[\protect\citeauthoryear{Chu, Weitz, Cheung, and Suciu}{Chu
  et~al\mbox{.}}{2017}]{ChuWCS17}
{\sc Chu, S.}, {\sc Weitz, K.}, {\sc Cheung, A.}, {\sc and} {\sc Suciu, D.}
  2017.
\newblock {HoTTSQL}: Proving query rewrites with univalent {SQL} semantics.
\newblock In {\em Proceedings of the 38th ACM SIGPLAN Conference on Programming
  Language Design and Implementation}. PLDI 2017. ACM, New York, NY, USA,
  510--524.

\bibitem[\protect\citeauthoryear{Clark}{Clark}{1977}]{adbt/Clark77}
{\sc Clark, K.~L.} 1977.
\newblock Negation as failure.
\newblock In {\em Logic and Data Bases}, {Gallaire} {and} {Minker}, Eds. Plenum
  Press, 293--322.

\bibitem[\protect\citeauthoryear{Cohen and Th\'ery}{Cohen and
  Th\'ery}{2017}]{Tarjan}
{\sc Cohen, C.} {\sc and} {\sc Th\'ery, L.} 2017.
\newblock Full script of {Tarjan SCC} {Coq}/{SSreflect} proof.
\newblock Tech. rep., INRIA.
\newblock \url{https://github.com/CohenCyril/tarjan} (visited: 2018-02).

\bibitem[\protect\citeauthoryear{??}{{Cypher}}{}]{Cypher}
{Cypher}.
\newblock \url{https://www.opencypher.org/} (visited: 2018-02).

\bibitem[\protect\citeauthoryear{Erling, Averbuch, Larriba-Pey, Chafi,
  Gubichev, Prat, Pham, and Boncz}{Erling et~al\mbox{.}}{2015}]{LDBC}
{\sc Erling, O.}, {\sc Averbuch, A.}, {\sc Larriba-Pey, J.}, {\sc Chafi, H.},
  {\sc Gubichev, A.}, {\sc Prat, A.}, {\sc Pham, M.-D.}, {\sc and} {\sc Boncz,
  P.} 2015.
\newblock The {LDBC} social network benchmark: Interactive workload.
\newblock In {\em Proceedings of the 2015 ACM SIGMOD International Conference
  on Management of Data}. SIGMOD '15. ACM, New York, NY, USA, 619--630.

\bibitem[\protect\citeauthoryear{Fan, Hu, and Tian}{Fan
  et~al\mbox{.}}{2017}]{FanHT17}
{\sc Fan, W.}, {\sc Hu, C.}, {\sc and} {\sc Tian, C.} 2017.
\newblock Incremental graph computations: Doable and undoable.
\newblock In {\em Proceedings of the 2017 ACM International Conference on
  Management of Data}. SIGMOD '17. ACM, New York, NY, USA, 155--169.

\bibitem[\protect\citeauthoryear{??}{{FlockDB}}{}]{FlockDB}
{FlockDB}.
\newblock \url{https://github.com/twitter-archive/flockdb} (visited: 2018-02).

\bibitem[\protect\citeauthoryear{??}{{G-Core}}{}]{gcore}
{G-Core}.
\newblock \url{https://github.com/ldbc/ldbc_gcore_parser} (visited: 2018-02).

\bibitem[\protect\citeauthoryear{??}{{Giraph}}{}]{Giraph}
{Giraph}.
\newblock \url{http://giraph.apache.org/} (visited: 2018-02).

\bibitem[\protect\citeauthoryear{Gonthier, Asperti, Avigad, Bertot, Cohen,
  Garillot, Roux, Mahboubi, O'Connor, Biha, Pasca, Rideau, Solovyev, Tassi, and
  Th{\'{e}}ry}{Gonthier et~al\mbox{.}}{2013}]{GonthierFT}
{\sc Gonthier, G.}, {\sc Asperti, A.}, {\sc Avigad, J.}, {\sc Bertot, Y.}, {\sc
  Cohen, C.}, {\sc Garillot, F.}, {\sc Roux, S.~L.}, {\sc Mahboubi, A.}, {\sc
  O'Connor, R.}, {\sc Biha, S.~O.}, {\sc Pasca, I.}, {\sc Rideau, L.}, {\sc
  Solovyev, A.}, {\sc Tassi, E.}, {\sc and} {\sc Th{\'{e}}ry, L.} 2013.
\newblock A machine-checked proof of the odd order theorem.
\newblock In {\em Interactive Theorem Proving}. {LNCS}. Springer Berlin
  Heidelberg, Berlin, Heidelberg, 163--179.

\bibitem[\protect\citeauthoryear{??}{{GraphQL}}{}]{GraphQL}
{GraphQL}.
\newblock \url{http://graphql.org/} (visited: 2018-02).

\bibitem[\protect\citeauthoryear{Greenlaw, Hoover, and Ruzzo}{Greenlaw
  et~al\mbox{.}}{1995}]{GreenLawHR95}
{\sc Greenlaw, R.}, {\sc Hoover, H.~J.}, {\sc and} {\sc Ruzzo, W.~L.} 1995.
\newblock {\em Limits to Parallel Computation: P-completeness Theory}.
\newblock Oxford University Press, Inc., New York, NY, USA.

\bibitem[\protect\citeauthoryear{??}{{Gremlin}}{}]{Gremlin}
{Gremlin}.
\newblock \url{http://tinkerpop.apache.org/} (visited: 2018-02).

\bibitem[\protect\citeauthoryear{Gupta, Mumick, and Subrahmanian}{Gupta
  et~al\mbox{.}}{1993}]{Gupta1993}
{\sc Gupta, A.}, {\sc Mumick, I.~S.}, {\sc and} {\sc Subrahmanian, V.~S.} 1993.
\newblock Maintaining views incrementally.
\newblock {\em SIGMOD Rec.\/}~{\em 22,\/}~2, 157--166.

\bibitem[\protect\citeauthoryear{Jagadish, Agrawal, and Ness}{Jagadish
  et~al\mbox{.}}{1987}]{Jagadish87}
{\sc Jagadish, H.~V.}, {\sc Agrawal, R.}, {\sc and} {\sc Ness, L.} 1987.
\newblock A study of transitive closure as a recursion mechanism.
\newblock {\em SIGMOD Rec.\/}~{\em 16,\/}~3, 331--344.

\bibitem[\protect\citeauthoryear{Leroy}{Leroy}{2009}]{DBLP:journals/cacm/Leroy09}
{\sc Leroy, X.} 2009.
\newblock Formal verification of a realistic compiler.
\newblock {\em Commun. {ACM}\/}~{\em 52,\/}~7, 107--115.

\bibitem[\protect\citeauthoryear{Letouzey}{Letouzey}{2008}]{Letouzey}
{\sc Letouzey, P.} 2008.
\newblock Extraction in {Coq}: An overview.
\newblock In {\em Proceedings of the 4th Conference on Computability in Europe:
  Logic and Theory of Algorithms}. CiE '08. Springer-Verlag, Berlin,
  Heidelberg, 359--369.

\bibitem[\protect\citeauthoryear{Mullen, Pernsteiner, Wilcox, Tatlock, and
  Grossman}{Mullen et~al\mbox{.}}{2018}]{Mullen:2018:MCE:3176245.3167089}
{\sc Mullen, E.}, {\sc Pernsteiner, S.}, {\sc Wilcox, J.~R.}, {\sc Tatlock,
  Z.}, {\sc and} {\sc Grossman, D.} 2018.
\newblock {\OE{}uf}: Minimizing the {Coq} extraction {TCB}.
\newblock In {\em Proceedings of the 7th ACM SIGPLAN International Conference
  on Certified Programs and Proofs}. CPP 2018. ACM, New York, NY, USA,
  172--185.

\bibitem[\protect\citeauthoryear{??}{{Neo4j}}{}]{Neo4j}
{Neo4j}.
\newblock \url{https://neo4j.com/} (visited: 2018-02).

\bibitem[\protect\citeauthoryear{??}{{Oracle PGX}}{}]{oraclepgx}
{Oracle PGX}.
\newblock
  \url{http://www.oracle.com/technetwork/oracle-labs/parallel-graph-analytix}
  (visited: 2018-02).

\bibitem[\protect\citeauthoryear{Ramalingam and Reps}{Ramalingam and
  Reps}{1996}]{RamalingamR96}
{\sc Ramalingam, G.} {\sc and} {\sc Reps, T.~W.} 1996.
\newblock On the computational complexity of dynamic graph problems.
\newblock {\em Theoretical Computer Science\/}~{\em 158,\/}~1{\&}2, 233--277.

\bibitem[\protect\citeauthoryear{Reutter, Romero, and Vardi}{Reutter
  et~al\mbox{.}}{2017}]{Reutter2017}
{\sc Reutter, J.~L.}, {\sc Romero, M.}, {\sc and} {\sc Vardi, M.~Y.} 2017.
\newblock Regular queries on graph databases.
\newblock {\em Theory of Computing Systems\/}~{\em 61,\/}~1, 31--83.

\bibitem[\protect\citeauthoryear{??}{{SPARQL}}{}]{SPARQL}
{SPARQL}.
\newblock \url{https://www.w3.org/TR/sparql11-query/} (visited: 2018-02).

\bibitem[\protect\citeauthoryear{{The Coq Development Team}}{{The Coq
  Development Team}}{2018}]{coqref}
{\sc {The Coq Development Team}}. 2018.
\newblock The {Coq} proof assistant, version 8.7.2.

\end{thebibliography}
\appendix



\section{Notations and Proofs Highlights}
\label{sec:notation}
The main notations used in the paper are summarized in
Table.~\ref{tab:NotationTable}.
\begin{table}[htbp]
\begin{center}
\begin{tabular}{r c l }
\hline
 $\gsig, \gvar, \gdom$  & $\triangleq$ & Symbol (Signature), Variable, and Constant (Domain) Sets \\
 $\sigma, \overline{\sigma}, \eta$ & $\triangleq$ & Substitution, Substitution Extension, and Closed Substitution (Grounding)\\
 $\strl, \strg$        	& $\triangleq$ & Already and To-Be Processed Stratas \\
 $\Pi, \g, \up, \uP, \uN$ & $\triangleq$ & \rd Program, Graph Instance, Batch Updates, Insertions, and Deletions \\
 $\uP(s), \uN(s)$       & $\triangleq$ & Batch Insertions and Deletions for symbol $s$ \\
 $V, V[\g]$             & $\triangleq$ & Top-Level \rd Program View, View Materialization over Base Instance \g \\
 $V[\appd{\g}{\up}]$    & $\triangleq$ & View Re-Materialization over Updated Instance $\appd{\g}{\up}$ \\
 $\Delta V[\appd{\g}{\up}]$	& $\triangleq$ & Incremental View Update \\
 $\iTp(\strl,\strg,\up)$    & $\triangleq$ & Program Maintenance Operator (\C|fwd_program|) \\
 $\iTBcl{s}(\g)$            & $\triangleq$ & Base Clausal Maintenance Operator (\C|fwd_or_clause_base|) \\
 $\iTcl{s}(\up)$        & $\triangleq$ & Incremental Clausal Maintenance Operator (\C|fwd_or_clause_delta|) \\

 $\matchb, \matchdb$    & $\triangleq$ & Base and Incremental Body Matching (\C|match_body, match_delta_body|)  \\
 $\matcha, \matchda$   & $\triangleq$ & Base and Incremental Atom Matching (\C|match_atom, match_delta_atom|)  \\
\hline
\end{tabular}
\end{center}
\caption{Notation Table}
\label{tab:NotationTable}
\end{table}
%

%
%
%
\label{sec:proof_details}
We describe in some more detail the theories used in our formal proof,
in particular we summarize the main lemma for \emph{modular} model
reasoning as well as the intermediate soundness results. The reader is
encouraged to look at the development directly, whose definitions are
intended to be readable and understood even by non-experts.

In the rest of the section, $\g$ is assumed to be a labelled graph,
$g$ a non-labelled graph (set of edges), $\up$ an update, $\Pi$ a
program, $C$ a clause, $\sset$, $\strl$, $\strg$ set of symbols, and
$s$ a symbol.

\subsection{Formal Theory for Modular Satisfaction}

\begin{lemma}[Modularity of Clause Satisfaction ({\inlinelink[\#sTc_mod]{VUP.vup}{\C|sTc_mod|}})]
  Assume $s \notin \sym(\up)$ and also
  $\sym(C) \cap \sym(\up) = \emptyset$. Then,
  $\dsat{s}{C} \iff \csat{s}{C}$.
\end{lemma}

\begin{lemma}[Modularity of Program Satisfaction ({\inlinelink[\#ssTp_mod]{VUP.vup}{\C|ssTp_mod|}})]
  Assume $\sset$ a \emph{well-formed slice} of $\Pi$ and
  $s \notin \sset$. Let $\up' = (\uP', \uN')$, where
  $\uP' = \uP \cup \{s(t_1, t_2) \mid (t_1, t_2) \in g \}$ and
  $\uN' = \uN \setminus \{s(t_1, t_2) \mid (t_1, t_2) \in g \}$.
  Then,

  \begin{equation*}
    \dsat[\up']{\{s\} \cup \sset}{\Pi} \iff \appd{\g}{\up'} \models_{s} \Pi(s) \wedge \appd{\g}{\up} \models_{\sset} \Pi
  \end{equation*}
\end{lemma}

\subsection{Formal Theory for Clause-level Operators}

\begin{lemma}[Soundness of Base Clausal Maintenance ({\inlinelink[\#fwd_or_clause_baseP]{VUP.vup}{\C|fwd_or_clause_baseP|}})]
  \label{lemma:fwdorclbP}
  \hfill Assume (H1) $\Pi(s)$ is a \emph{safe} clause; (H2) $\strl$ is
  complete for closures; (H3) $\strl$ is a \emph{well-formed slice} of
  $\Pi$; (H4) $s \notin \strl$; (H) $\sym(\Pi(s)) \subseteq \strl$,
  and $\csat{\strl}{\Pi}$. Then $\dsat[\up_s]{\uset{s}{\strl}}{\Pi}$,
  where $\up_{s} = \iTBcl{s}(\appd{\g}{\up})$.
 \end{lemma}

\begin{lemma}[Soundness of Incr. Clausal Maintenance ({\inlinelink[\#fwd_or_clause_deltaP]{VUP.vup}{\C|fwd_or_clause_deltaP|}})]
  \label{lemma:fwdorcldP}
  Assume (H1) $\Pi(s)$ is a \emph{safe} clause; (H2) $\strl$ is
  complete for closures; (H3) $\strl$ is a \emph{well-formed slice} of
  $\Pi$; (H4) $s \notin \strl$; (H5) $\sym(\Pi(s)) \subseteq \strl$;
  (H6) $\dsat{\strl}{\Pi}$; (H7) $\sym(\up) \subseteq \strl$.  Also
  assume the \emph{incrementality conditions}: (H8)
  $\csat{\sset}{\Pi}$; (H9) $s \in \sset$; (H10)
  $\sym(\Pi(s)) \cap \sym(\uN)$.  Then,
  $\dsat[\up_s]{\uset{s}{\strl}}{\Pi}$, where
  $\up_{s} = T_{\g}^{\Pi,s}(\up)$. This operator is called by
  the supported maintenance operator when incrementality can be used.
\end{lemma}

\begin{lemma}[Soundness of Supported Clausal Maintenance ({\inlinelink[\#fwd_or_clauseP]{VUP.vup}{\C|fwd_or_clauseP|}})]
  \label{lemma:fwdorclP}
  \hfill Assume (H1) $\Pi(s)$ is a \emph{safe} clause, (H2)
  $\csat{\sset}{\Pi}$; (H3) $\strl$ is well-formed wrt closures; (H4)
  $\strl$ is a \emph{well-formed slice} of $\Pi$; (H5)
  $s \notin \strl$; (H6) $\sym(\Pi(s)) \subseteq \strl$; (H7)
  $\sym(\up) \subseteq \strl$. If $\dsat{\strl}{\Pi}$, then
  $\dsat[\up_s]{\{s\} \cup \strl}{\Pi}$, where
  $\up_{s} = \iTcl{s}(\up)$.
\end{lemma}

\begin{lemma}[Soundness of Incr. Body Matching ({\inlinelink[\#fwd_delta_body_sound]{VUP.vup}{\C|fwd_delta_body_sound|}})]
  \label{lemma:cbm}
  Let $B$ a conjunctive body; $\sigma$ a substitution. Assume
  $\sym(B) \cap \sym(\uN) = \emptyset$, that is to say, no deletions
  are scheduled for $B$, then for all $\sigma \in \matchdb(B)$ there
  exists an instantiation of $B$, $\overline{B}$, such that
  $\sigma(B) = \overline{B}$.
\end{lemma}

\end{document}